\documentclass[11pt,letterpaper]{article}

\usepackage{jheppub}
\usepackage{color}
\usepackage{graphicx}
\usepackage{wrapfig}

\usepackage{verbatim}
\usepackage{amsmath}
\usepackage{amssymb}
\usepackage{subfig}
\usepackage{url}
\usepackage{bbold}
\usepackage{xspace}

\usepackage{multirow}
\usepackage{threeparttable}
\usepackage{paralist}

\usepackage{color}

\usepackage{floatrow}
\newfloatcommand{capbtabbox}{table}[][\FBwidth]

\usepackage{slashed}
\usepackage{verbatim}

\newcommand{\met}{E\!\!\!/_T}

\usepackage{tikz}
\usetikzlibrary{trees}
\usetikzlibrary{decorations.pathmorphing}
\usetikzlibrary{decorations.markings}
\usetikzlibrary{arrows}

\definecolor{darkgreen}{rgb}{0,0.5,0}

\newcommand{\be}{\begin{equation}}
\newcommand{\ee}{\end{equation}}

\DeclareRobustCommand{\Fig}[1]{Fig.~\ref{#1}}

\DeclareRobustCommand{\Eq}[1]{Eq.~(\ref{#1})}

\newcommand{\Dfbd}{\mathord{\buildrel{\lower3pt\hbox{$\scriptscriptstyle\leftrightarrow$}}\over {D}_{\mu}}}

\newcommand{\beq}{\begin{equation}}
\newcommand{\eeq}[1]{\label{#1}\end{equation}}
\def\beqa{\begin{eqnarray}}
\def\eeqa#1{\label{#1}\end{eqnarray}}
\newcommand{\eeqn}{\end{equation}}

\def\stacksymbols #1#2#3#4{\def\theguybelow{#2}
    \def\vp{\lower#3pt}
    \def\sp{\baselineskip0pt\lineskip#4pt}
    \mathrel{\mathpalette\intermediary#1}}

\def\intermediary#1#2{\vp\vbox{\sp
     \everycr={}\tabskip0pt
     \halign{$\mathsurround0pt#1\hfil##\hfil$\crcr#2\crcr
              \theguybelow\crcr}}}

\begin{document}

\tikzset{
	  photon/.style={decorate, decoration={snake}, draw=black},
	  boson/.style={decorate, decoration={snake}, draw=black},
	  electron/.style={draw=black, postaction={decorate},
	           decoration={markings,mark=at position .55 with {\arrow[draw=black]{latex}}}
	  },
	  electron2/.style={draw=black, postaction={decorate},
	           decoration={markings,mark=at position .55 with {\arrow[draw=black]{latex reversed}}}
	  },
	  fermion/.style={draw=black, postaction={decorate},
	            decoration={markings,mark=at position .55 with {\arrow[draw=black]{}}}
	  },
 pttt/.style={decorate, draw=white},
	  gluon/.style={decorate, draw=black, 
	    decoration={coil,amplitude=4pt, segment length=6pt}},
	  gluon/.style={decorate, draw=black, 
	    decoration={coil,amplitude=4pt, segment length=6pt}},
	  higgs/.style={draw=black, postaction={decorate},
	           decoration={markings,mark=at position .55 with}
	  },
	  nothing/.style={draw=white}
	}

\title{The Higgs Portal Above Threshold}

\author[a]{Nathaniel Craig,}
\author[b]{Hou Keong Lou,}
\author[c]{Matthew McCullough,}
\author[d]{and Arun Thalapillil}

\affiliation[a]{Department of Physics, University of California, Santa Barbara, CA 93106, USA}
\affiliation[b]{Department of Physics, Princeton University, Princeton, NJ 08540, USA}
\affiliation[c]{Theory Division, CERN, 1211 Geneva 23, Switzerland}
\affiliation[d]{Department of Physics and Astronomy, Rutgers University, Piscataway, NJ 08854, USA}

\emailAdd{ncraig@physics.ucsb.edu}
\emailAdd{hlou@princeton.edu}
\emailAdd{matthew.mccullough@cern.ch}
\emailAdd{thalapillil@physics.rutgers.edu}

\date{\today}

\abstract{The discovery of the Higgs boson opens the door to new physics interacting via the Higgs Portal, including motivated scenarios relating to baryogenesis, dark matter, and electroweak naturalness. We systematically explore the collider signatures of singlet scalars produced via the Higgs Portal at the 14 TeV LHC and a prospective 100 TeV hadron collider. We focus on the challenging regime where the scalars are too heavy to be produced in the decays of an on-shell Higgs boson, and instead are produced primarily via an off-shell Higgs. Assuming these scalars escape the detector, promising channels include missing energy in association with vector boson fusion, monojets, and top pairs. We forecast the sensitivity of searches in these channels at $\sqrt{s} = 14$ \& 100 TeV and compare collider reach to the motivated parameter space of singlet-assisted electroweak baryogenesis, Higgs Portal dark matter, and neutral naturalness.}


\arxivnumber{14xx.xxxx}

\preprint{}

\maketitle

\section{Introduction}

The discovery of the Higgs boson \cite{Aad:2012tfa, Chatrchyan:2012ufa} provides unprecedented opportunities in the search for physics beyond the Standard Model (SM). More than any other state in the Standard Model, the Higgs is a sensitive barometer of new physics. Perhaps the most familiar opportunity involves Higgs couplings; the rigidity of electroweak symmetry breaking in the Standard Model uniquely determines the interactions of the SM Higgs, such that any deviations in couplings would be an unambiguous indication of new physics. But the Higgs also provides an entirely new gateway to physics beyond the Standard Model thanks to the low dimension of the operator $|H|^2$: it admits new marginal or relevant operators of the form $|H|^2 \mathcal{O}$, where $\mathcal{O}$ is a gauge-invariant operator with $\Delta_{\mathcal{O}} \lesssim 2$. The classic example is $\mathcal{O} = \phi^2$ were $\phi$ is neutral under the SM but enjoys a $Z_2$ symmetry \cite{Silveira:1985rk, McDonald:1993ex, Burgess:2000yq, Patt:2006fw, Barger:2007im}. This {\it Higgs Portal} provides an entirely new avenue to access physics beyond the Standard Model. Such portals are motivated not only on purely pragmatic grounds as one of only two possible marginal couplings between the SM and SM-singlet states, but also on theoretical grounds in diverse scenarios relating to dark matter, electroweak baryogenesis, and solutions to the gauge hierarchy problem. Now that the Higgs boson has been discovered, the exploration of possible Higgs Portals and their signatures has become a high priority at the LHC and future colliders.

In this paper we consider the scalar Higgs Portal consisting of
\begin{equation} \label{eq:portal}
\mathcal{L} = \mathcal{L}_{SM} - \frac{1}{2} \partial_\mu \phi \partial^\mu \phi - \frac{1}{2} M^2 \phi^2 - c_\phi |H|^2 \phi^2
\end{equation}
where $H$ is the SM-like Higgs doublet and $\phi$ is a scalar neutral under the Standard Model.\footnote{In this work we will neglect other ``portal'' couplings to fermions or vector bosons neutral under the Standard Model. Such interactions are irrelevant (and in the case of vector bosons, not even gauge invariant) and should often be properly treated by integrating in additional states.}   We have taken $\phi$ to be a real scalar, but it could equally well be a complex scalar and carry charges under additional gauge sectors. The coupling $c_\phi$ can take arbitrary values, but in Higgs Portals motivated by baryogenesis or naturalness, $c_\phi$ is often $\mathcal{O}(1)$. The $\phi$ field may also possess self-couplings relevant for baryogenesis or couplings to additional states in the hidden sector, but these are in general irrelevant to determining how well the portal coupling of (\ref{eq:portal}) can be probed directly to discover or exclude the scalar $\phi$.

There are many cases in which $\phi$ is relatively easy to detect. If $\phi$ acquires a vacuum expectation value then Higgs-singlet mixing can leave direct signals in Higgs couplings and the production and decay of a heavy Higgs state \cite{Barger:2007im}, both of which may be probed effectively at the LHC and future $e^+ e^-$ machines. Far more challenging is the scenario where the $\phi$ respects an unbroken $Z_2$ symmetry, $\phi \to - \phi$, in which case there is no Higgs-singlet mixing and the couplings of the Higgs are unaltered at tree level. After electroweak symmetry breaking the theory consists of
\begin{equation} \label{eq:brokenportal}
\mathcal{L} = \mathcal{L}_{SM} - \frac{1}{2} \partial_\mu \phi \partial^\mu \phi - \frac{1}{2} m_\phi^2 \phi^2 - c_\phi v h \phi^2 - \frac{1}{2} c_\phi h^2 \phi^2
\end{equation}
where $m_\phi^2 = M^2 + c_\phi v^2$ in units where $v = 246$ GeV. The Higgs Portal coupling is the only connection between $\phi$ and the Standard Model, and so the only available production mode at colliders is $\phi \phi$ production via the Higgs boson. By assumption $\phi$ has no SM decay modes and appears as missing energy in collider detectors. Discovering or excluding such a Higgs Portal at $pp$ machines requires focusing on Higgs associated production modes in order to identify the missing energy signal.\footnote{Throughout we will take $c_\phi$ as a free parameter with values up to rough perturbative bounds at the relevant scale.  However we do not consider RG effects which may also be interesting for constraining large couplings by requiring the absence of Landau poles, as these depend sensitively on additional dynamics in the hidden sector.}

When $m_\phi < m_h/2$ this scenario may be very efficiently probed via the Higgs invisible width \cite{Djouadi:2011aa, Englert:2011us, Djouadi:2012zc, Englert:2013gz, Aad:2014iia,Chatrchyan:2014tja}, since the Higgs can decay on-shell into $\phi$ pairs and the smallness of the SM Higgs width ensures the rate for $pp \to h + X \to \phi \phi + X$ is large for a wide range of $c_\phi$. When $m_\phi > m_h/2$, however, the Higgs cannot decay on-shell to $\phi \phi$, and $\phi$ pair production instead proceeds through an off-shell Higgs, $pp \to h^* + X \to \phi \phi + X$. The cross section for this process is then suppressed by an additional factor of $|c_\phi|^2$ as well as two-body phase space, leading to a rapidly diminishing rate and extremely challenging prospects at the LHC. Nonetheless, this may be the only avenue for discovering or excluding Higgs Portals above the kinematic threshold for production via an on-shell Higgs boson. 

In this paper we seek to determine the prospects for exclusion or discovery of $\phi$ at the LHC and future colliders when $m_\phi > m_h/2$. For simplicity we focus on $\sqrt{s} = 14$ TeV at the LHC and $\sqrt{s} = 100$ TeV at a future $pp$ collider; the sensitivity at lepton colliders was studied extensively in \cite{Chacko:2013lna}. Although the possibility of probing Higgs Portal states via an off-shell Higgs was identified even before Higgs discovery \cite{Djouadi:2011aa, Djouadi:2012zc}, collider studies to date have been somewhat limited. In \cite{Endo:2014cca} the state of existing limits was established by reinterpretation of LHC searches for invisible Higgs decays at $\sqrt{s} = 8$ TeV in terms of vector boson fusion, gluon associated production, and $Z$ associated production via an off-shell Higgs, along with limited projections for $\sqrt{s} = 14$ TeV. Sensitivity to the novel $H$-Higgsstrahlung mode has been studied at $\sqrt{s} = 8$ \& 14 TeV  \cite{Carpenter:2013xra}, while sensitivity at $\sqrt{s} = 100$ TeV has been broached in a limited study of vector boson fusion production \cite{Curtin:2014jma}. Further study at $\sqrt{s} = 14$ \& 100 TeV is strongly motivated, both to help optimize future searches at the LHC and to establish the physics case for a future hadron collider.
 
The most promising channel at $pp$ colliders is vector boson fusion (VBF) production of $\phi$ pairs via an off-shell Higgs boson, leading to a signal with two forward jets and missing energy. Ancillary channels sensitive to the missing energy signal include gluon fusion production (ggH) with an associated jet, $t \bar t$ associated production (ttH), $Z$-Higgsstrahlung (ZH), and the novel $H$-Higgsstrahlung (HH) channel giving rise to mono-Higgs plus missing energy \cite{Carpenter:2013xra}. Each has relative virtues. The cross section for ggH production is largest at $\sqrt{s} = 14, 100$ TeV but the additional jet requirement and kinematic separation of signal from background reduces signal significance. The ttH cross section is significantly smaller but grows substantially at 100 TeV, and the $t \bar t + \met$ final state has already proven sensitive to invisible decays of the Higgs boson at 8 TeV \cite{Zhou:2014dba}. The cross section for ZH production is among the smallest of the modes and not well separated from the $Z +\nu\nu$ backgrounds, rendering it less promising. The $h + \met$ signature of HH production is particularly interesting, as it directly probes the Higgs Portal interaction, but preliminary study at $\sqrt{s} = 8, 14$ TeV  \cite{Carpenter:2013xra} suggests far less sensitivity than the VBF mode \cite{Endo:2014cca}.\footnote{The interpretation of \cite{Carpenter:2013xra} for $c_\phi > 1, m_\phi \lesssim v$ is also unclear, as in this regime the mono-Higgs final state accumulates comparable contributions from both single and double insertions of the Higgs Portal interaction.} Consequently, the balance of production cross section and background separation provided by VBF render it the most promising of the channels, but for completeness in this work we will consider the prospects of VBF, ggH, and ttH searches at $\sqrt{s} = 14$ and 100 TeV.

Note there is also a complementary, indirect means of probing this scenario through its impact on precision Higgs coupling measurements. The interaction (\ref{eq:portal}) leads to shifts in the Higgs-$Z$ coupling relative to the Standard Model that may be probed through measurements of the $Zh$ production cross section at future $e^+ e^-$ colliders \cite{Englert:2013tya, Craig:2013xia}. Precision on $\delta \sigma_{Zh}$ is expected to approach the level of $\sim 0.32 \%$ at $1\sigma$ at circular $e^+ e^-$ colliders such as CEPC/TLEP \cite{Dawson:2013bba}. A particularly interesting question is whether significant deviations in $\sigma_{Zh}$ at an $e^+ e^-$ collider may be followed by conclusive evidence for (\ref{eq:portal}) at a future $pp$ collider.

Our paper is organized as follows: We begin in Section \ref{sec:np} by reviewing three motivated scenarios for physics beyond the Standard Model giving rise to Higgs Portal interactions. This helps to motivate regions of the Higgs Portal parameter space that might be probed by direct searches. In Section \ref{sec:search} we outline our procedure for simulating Higgs Portal searches at the LHC and future colliders in vector boson fusion, gluon associated production, and $t \bar t$ associated production. In Section \ref{sec:discussion} we present the exclusion and discovery reach for searches at $\sqrt{s} = 14$ TeV and 100 TeV and discuss the implications for motivated new physics scenarios. We conclude in Section \ref{sec:conclusions} and reserve some of the details of our electroweak baryogenesis parameterization for Appendix \ref{app:ewbg}.

\section{New physics through the Higgs Portal} \label{sec:np}

Although the Higgs Portal is motivated purely as a leading operator through which generic new physics might couple to the Standard Model, there are a variety of specific scenarios for physics beyond the Standard Model (BSM) in which Higgs Portal couplings feature prominently. These scenarios provide a motivated range of masses and couplings against which we can benchmark the reach of Higgs Portal searches at the LHC and future colliders.

\subsection{Electroweak baryogenesis}

A particularly motivated scenario for Higgs Portal interactions arises in models of baryogenesis aimed at generating the observed asymmetry between baryons and anti-baryons. The Standard Model famously contains the ingredients necessary for baryogenesis to occur {\it in principle} during the electroweak phase transition, realizing the scenario of electroweak baryogenesis (EWBG). However, while the ingredients are present for electroweak baryogenesis to occur in principle, in practice the parameters of the Standard Model are such that the electroweak phase transition is too weak to realize the necessary departure from equilibrium \cite{Kuzmin:1985mm}. The phase transition may be rendered sufficiently first-order if the Higgs couples strongly to additional light degrees of freedom, potentially connecting the Higgs Portal to EWBG.

The general story of electroweak baryogenesis is well known. In the early universe, one expects electroweak symmetry to be restored at high temperature \cite{Weinberg:1974hy,Quiros:1999jp}. The net baryon number is zero as any deviation will be washed out by electroweak sphalerons, which are unsuppressed in the unbroken phase. As the temperature cools to near the critical temperature $T_c$, the unbroken and broken phase become roughly degenerate. Bubbles of broken electroweak symmetry begin to form. With sufficient CP violation \cite{Dine:1990fj}, interactions with surrounding particles will produce a net baryon number \cite{ Anderson:1991zb, Dine:1992wr, Cohen:1993nk, Rubakov:1996vz,  Trodden:1998ym}. Within the bubble, electroweak symmetry is broken and sphaleron transition rates are highly suppressed, such that the generated baryon asymmetry is maintained. This requires the Boltzmann suppression for the sphaleron to be sufficiently high \cite{Patel:2011th}
\begin{equation}
e^{-\Delta E_{\rm sphaleron}/T_c} \lesssim e^{-10}
\end{equation}
Since it is generally difficult to compute the sphaleron energy, one typically approximates the baryon asymmetry condition by computing the thermal potential and demanding \cite{Quiros:1999jp}
\begin{equation}
\frac{\phi(T_c)}{T_c} \gtrsim 1.0
\label{eq:ewbg}
\end{equation}
where $\phi(T_c) \equiv v_c$ is the vacuum expectation of the Higgs field in the broken phase at the critical temperature. Despite the fact that equation (\ref{eq:ewbg}) is gauge dependent, we will use it as a leading order estimate, since even a careful result will still require two-loop calculations in order to be reliable ~\cite{Patel:2011th}.

In the Standard Model itself, the electroweak phase transition is not strong enough to satisfy the condition (\ref{eq:ewbg}), requiring BSM states \cite{Espinosa:1993bs, Carena:2004ha, Ham:2005ej, Profumo:2007wc, Carena:2008vj, Cohen:2011ap, Curtin:2012aa} or corrections to the SM EFT \cite{Grojean:2004xa} to alter the Standard Model effective potential and render the phase transition strongly first-order. To properly influence the effective potential, the new states must be light (in order to be relevant during the electroweak phase transition) and relatively strongly coupled to the Higgs boson (in order to substantially alter the thermal potential). This raises the tantalizing prospect of discovering or falsifying electroweak baryogenesis through direct searches at colliders. If the new states possess Standard Model quantum numbers, they may be readily detected through either direct searches or indirect effects on Higgs couplings \cite{Profumo:2014opa, Katz:2014bha}. But singlet scalars coupling through the Higgs portal are also a viable candidate, with correspondingly weaker prospects for direct and indirect probes. In this respect it is particularly worthwhile to study the sensitivity of the LHC and future colliders to singlet-assisted electroweak baryogenesis.

The parameter space of singlet-assisted electroweak baryogenesis was recently considered in detail in \cite{Curtin:2014jma} (see also \cite{Espinosa:2007qk, Barger:2007im,  Noble:2007kk, Espinosa:2011ax, Cline:2012hg, Chung:2012vg, Cline:2013gha, Profumo:2014opa, Fuyuto:2014yia}), and we will largely follow their discussion here. There are two possibilities for singlet-assisted electroweak baryogenesis, corresponding to a single-step and a double-step phase transition, respectively. The single-step phase transition proceeds purely along the Higgs direction, where the role of the singlet(s) is to correct the Higgs effective potential to render the electroweak phase transition strongly first-order. Alternately, if $M^2 < 0$, the universe can first undergo a transition to a vacuum along the singlet direction and then proceed to the EWSB vacuum. This amounts to a tree-level modification of the Higgs potential and can lead to an arbitrarily strong first-order phase transition. At the level of the Higgs Portal model in (\ref{eq:portal}), $M^2 < 0$ corresponds to a runaway direction, but this may be stabilized by a quartic coupling of the form $\lambda \phi^4$ and the strength of the phase transition dialed by adjusting $\lambda$. A third possibility is for a one-step phase transition to proceed via thermal effects as in the MSSM, but this occurs strictly in the two-step regime. 

\begin{figure}[tbp] 
   \centering
 \includegraphics[width=5.0in]{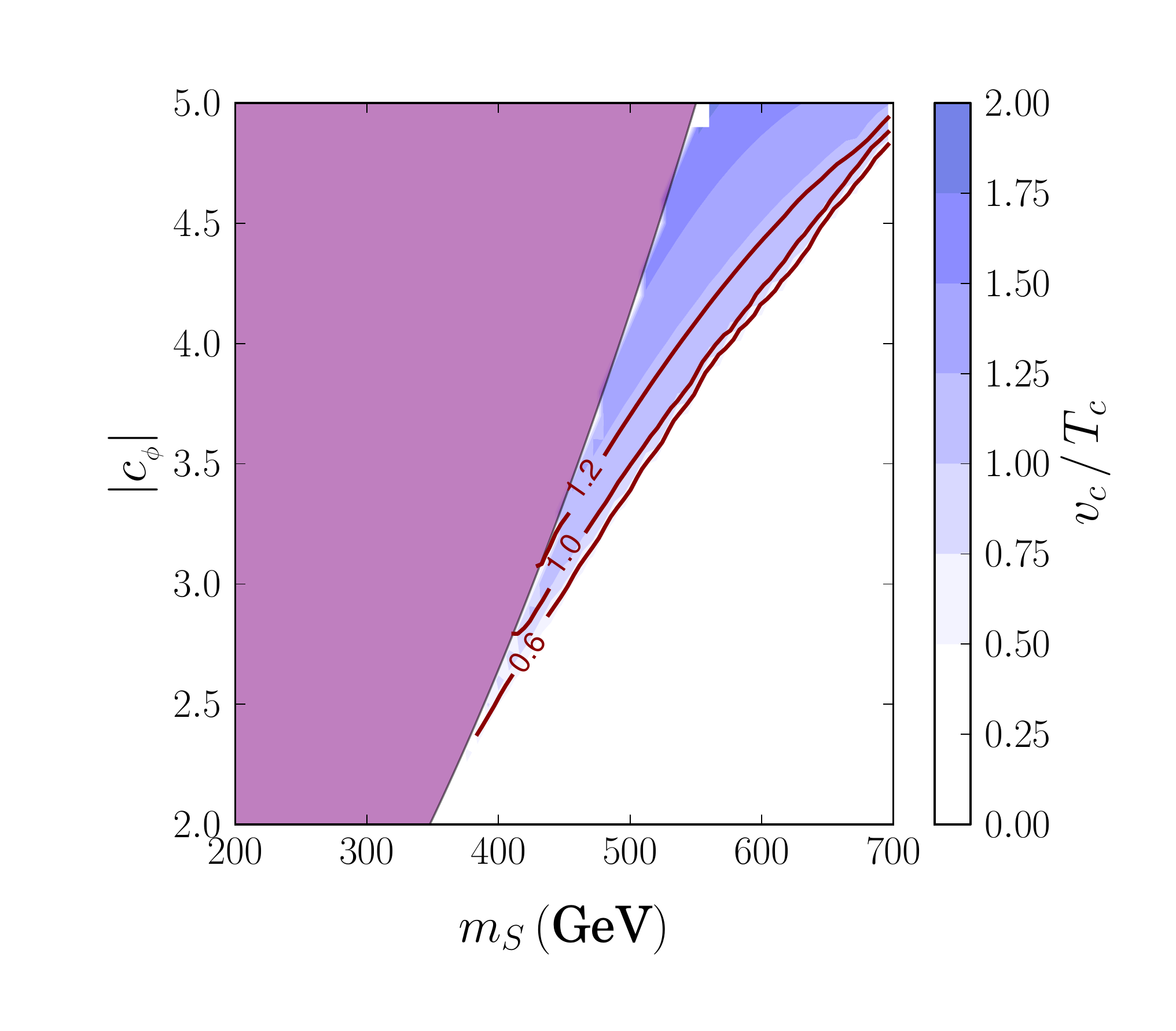}
{   \caption{Values of $v_c/T_c$ as a function of $m_\phi$ and $c_\phi$ for singlet-assisted electroweak baryogenesis. The shaded purple region indicates a runaway for the Higgs Portal model corresponding to $M^2 < 0$, but the runaway can be stabilized by adding a quartic coupling of the form $\lambda \phi^4$.  A perturbative two-step phase transition may then proceed in a $\sim 50$ GeV-wide penumbra at the edge of the shaded purple region, while a two-step transition deeper in the purple region cannot be ruled out but requires nonperturbatively large values of $\lambda$ to ensure the EWSB vacuum is deeper than the singlet vacuum \cite{Curtin:2014jma}. }
   \label{fig:ewbg}}
\end{figure}

We illustrate the viable parameter space of EWBG in the Higgs Portal model in Fig.~\ref{fig:ewbg}. A two-step transition may occur in the region corresponding to $M^2 < 0$ (i.e., $c_\phi v^2 > m_\phi^2$), with the proper ordering of the singlet vacuum and EWPT vacuum achieved by dialing the quartic $\lambda$. For modest negative values of $M^2$ this two-step transition is under perturbative control, but far in the $M^2 < 0$ region this requires nonperturbatively large $\lambda$ where we lack control but cannot definitely rule out EWBG. In the region with $M^2 > 0$ we plot contours of $v_c / T_c$ as a function of $c_\phi$ and $m_\phi$, allowing that EWBG may occur in regions with $v_c / T_c \gtrsim 0.6$ given unknown details of baryogenesis during the phase transition. We reserve some details of our calculation in this region for Appendix \ref{app:ewbg}. Our results for this region are in good agreement with the results presented in  \cite{Curtin:2014jma}. This provides a strongly motivated target for direct searches for Higgs Portal states at the LHC and future colliders.

\subsection{Dark matter}
\begin{figure}[tbp] 
   \centering
 \includegraphics[width=4.0in]{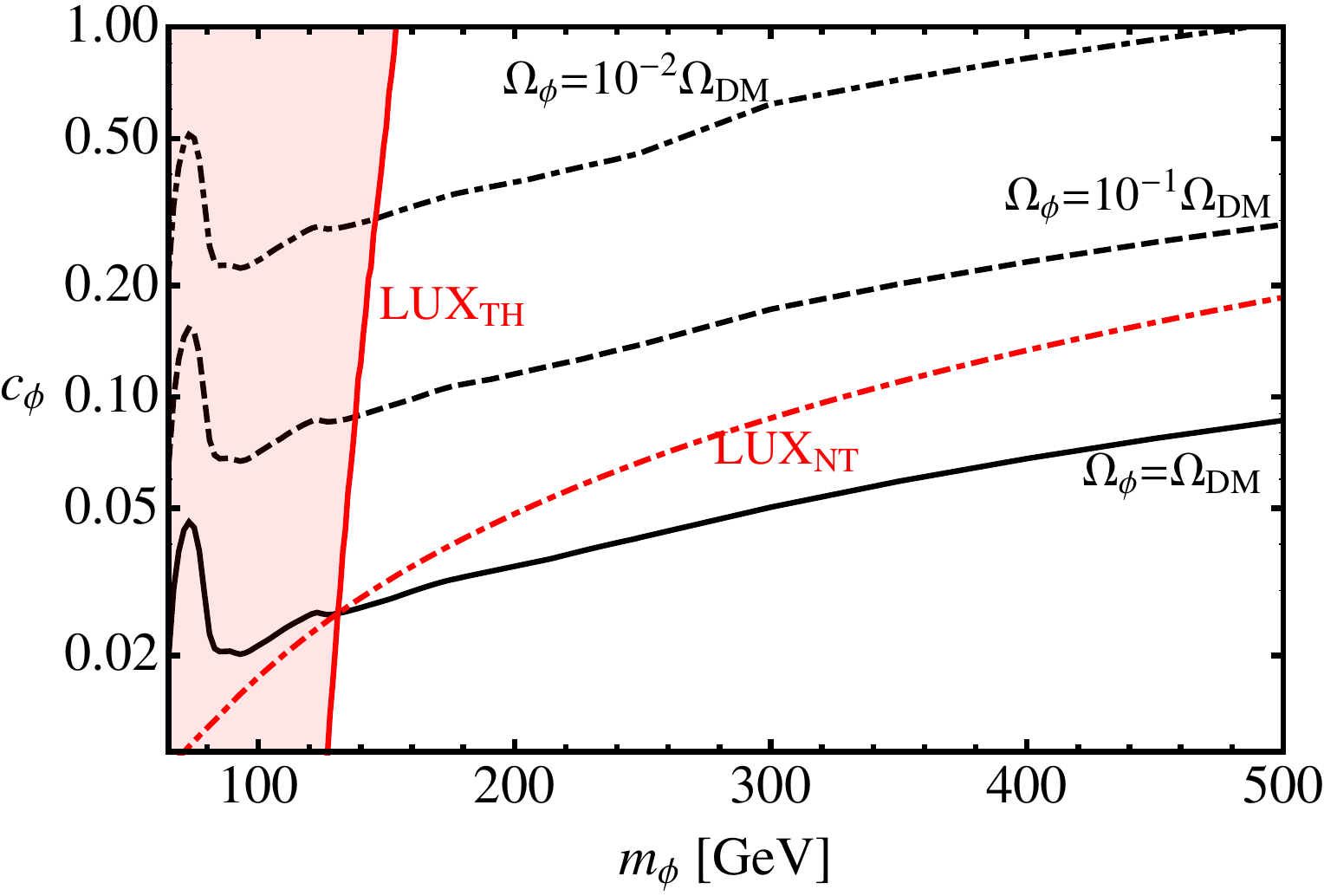}
  { \caption{Contours of relic DM density from freeze-out through the Higgs Portal.  Constraints on the parameter space from the LUX direct detection experiment \cite{Akerib:2013tjd} are shown in dotdashed red (labelled $\text{LUX}_{\text{NT}}$) where an additional assumption is made that in regions where thermal freeze-out over- or under-produces DM, additional fields and couplings lead to late-time DM dilution or production to realize the observed density.  On the other hand, the solid red line (labelled $\text{LUX}_{\text{TH}}$) and shaded region show the parameter space which is excluded if one only takes the Lagrangian of \Eq{eq:portal} and the assumption of a standard thermal history.  In this case $\Omega_\phi \propto c_\phi^{-2}$ and $\sigma_n  \propto c_\phi^{2}$ thus the exclusion is almost independent of the coupling, and largely depends only on the mass.}
   \label{fig:thermDM}}
\end{figure}

Throughout this work we assume that $\phi$ is charged under an approximate $Z_2$ such that it is stable on collider timescales.  However if the $Z_2$ symmetry is exact the Higgs Portal furnishes a dark matter candidate \cite{Silveira:1985rk,McDonald:1993ex,Burgess:2000yq,Davoudiasl:2004be,Patt:2006fw}, adding further motivation to collider searches for the Higgs Portal above threshold.  Higgs Portal dark matter is very predictive in the sense that if $\phi$ is required to provide the entirety of the observed dark matter abundance and this abundance arises thermally, then for a given $m_\phi$ the required coupling $c_\phi$ is determined.  This is shown in \Fig{fig:thermDM}, where it can be seen that requiring the observed DM abundance leads to relatively small couplings.\footnote{This relic abundance has been calculated using the formulae of \cite{Cline:2013gha}.}  If $\phi$ only accounts for some fraction of the dark matter, or if it is produced non-thermally in the early Universe from e.g.\ late decays of some other field, then this requires larger couplings.

Although it only communicates with the SM via the Higgs sector, current direct detection experiments are already sensitive to Higgs Portal dark matter.  In the dashed line of \Fig{fig:thermDM} the current bounds on $c_\phi$ from the LUX experiment \cite{Akerib:2013tjd} are shown assuming that $\phi$ comprises the entirety of the observed DM abundance.  Such a scenario typically requires either late-time dilution of the DM abundance to ameliorate over-production due to small couplings, or alternatively late-time DM production to counteract the over-annihilation of DM due to large couplings.\footnote{We have used the effective Higgs nucleon coupling $f_N=0.29$ as found in \cite{Junnarkar:2013ac,Agrawal:2014oha}.}  However, if only a standard thermal history is assumed with no entropy release or DM production below temperatures $T\sim m_\phi/20$ then regions where $\phi$ under-annihilates and is overproduced are excluded by observations.  In regions where it over-annihilates and comprises only some subcomponent of the DM the direct detection constraints must be re-weighted to account for the reduced abundance and it must be assumed that some other field makes up the total of the DM abundance.  In this case, with only the assumption of a standard thermal history, the direct detection constraints become weaker, as shown in the solid line of \Fig{fig:thermDM}.

\begin{figure}[tbp] 
   \centering
 \includegraphics[width=4.0in]{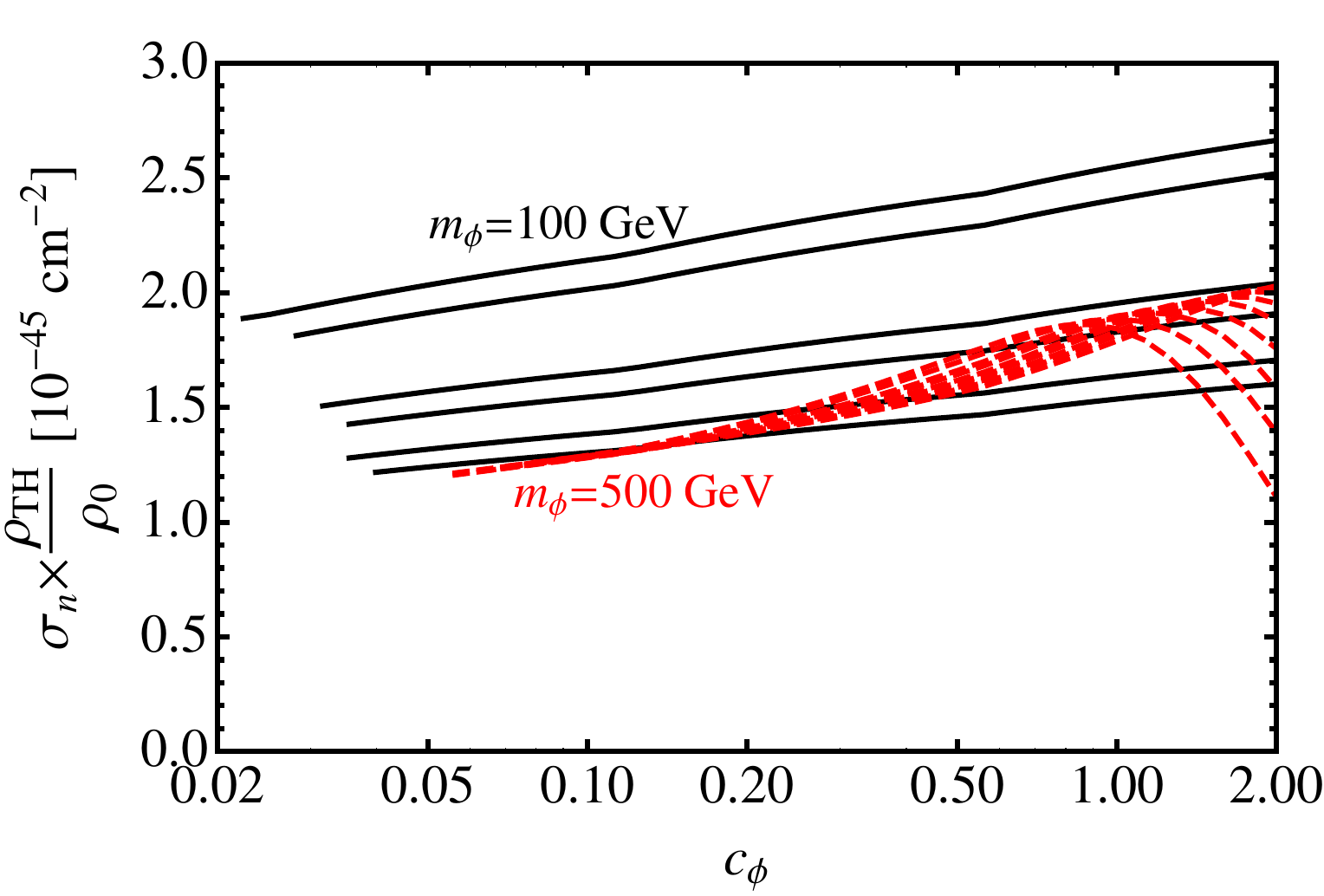}
  { \caption{The DM-nucleon scattering cross section weighted by the fractional relic density predicted by a standard thermal history $R_{DD} = \sigma_n \times \rho_{TH}/\rho_0$ as a function of the Higgs portal coupling $c_\phi$ for a variety of masses from $100 \text{ GeV} < m_\phi < 500 \text{ GeV} $ in steps of $25$ GeV.  Masses $m_\phi < 2 m_h$ are shown in solid black and $m_\phi > 2 m_h$ in dashed red.  Regions where DM is over-produced are not shown.  This demonstrates that direct detection predictions for the Higgs portal with a standard thermal history only depend very weakly on the coupling and exclusions stronger than $\sigma_n < \mathcal{O}(1 \times 10^{-45} \text{cm}^2)$ typically exclude the Higgs portal with a standard thermal history independent of the coupling.  The suppression with large coupling is shown whenever the $\phi + \phi \to h + h$ starts to dominate when kinematically accessible.}
  \label{fig:DD}}
\end{figure}

The complementarity of direct detection and collider probes of Higgs Portal DM can be understood from some simple scaling arguments.  For $m_\phi < m_h$ the DM annihilates through an s-channel Higgs.  Thus $\langle \sigma v \rangle \propto c_\phi^2$ and $\Omega_\phi h^2 \propto c_\phi^{-2}$.  The direct detection cross section scales as $\sigma_n \propto c_\phi^2$.  Taking the product of this cross section with the relic density and assuming a standard thermal history in this mass range leads to overall direct detection rates ($R_{DD} = \sigma_n \times \rho_{TH}/\rho_0$) which are largely insensitive to the coupling $R_{DD} \propto c_\phi^0$.  Interestingly this implies that for the Higgs Portal with a thermal history, predicted direct detection rates in this mass range are almost independent of the Higgs Portal coupling, and the predicted rate essentially becomes a function of the mass only.  This is demonstrated in \Fig{fig:DD} where the cross-section is weighted by the fractional density of Higgs Portal DM from a standard thermal history to give $R_{DD}$ as a function of $c_\phi$.  The coupling $c_\phi$ is varied over two orders in magnitude, however the direct detection rate predicted by a standard thermal history only varies by $\mathcal{O}(10 \text{'s} \%)$.  This demonstrates that over the mass range $100 \text{ GeV} < m_\phi < 500 \text{ GeV} $ direct detection exclusions stronger than $\mathcal{O}(1 \times 10^{-45} \text{cm}^2)$ actually exclude the Higgs portal with a standard thermal history independent of the Higgs portal coupling.  It should be noted that non-standard thermal histories may significantly modify the constraint.

For $m_\phi > m_h$ annihilation can also proceed into two Higgs bosons, hence for large $c_\phi$ we have $\langle \sigma v \rangle \propto c_\phi^4$ in the limit of large $c_\phi$.  Now taking the product of thermal abundance and direct detection cross section we have $R_{DD} \propto c_\phi^{-2}$.  Thus the suppression of relic density can overcome the enhancement of scattering cross section and a standard thermal history leads to smaller direct detection rates for larger couplings, as demonstrated in \Fig{fig:DD} .  On the other hand, the collider cross sections scale as $c_\phi^2$, with the exception of mono-Higgs signals which scale as a polynomial up to $c_\phi^4$.

Thus we are led to a strong sense of complementarity between direct detection and collider probes of the Higgs Portal.
\begin{itemize}
\item  If the $Z_2$ symmetry is exact and a saturation of the observed DM density is assumed (which may require a non-thermal history), then direct detection probes are likely to be most sensitive.
\item  If the $Z_2$ symmetry is exact and a standard thermal history is assumed then in regions where $\Omega_\phi \leq \Omega_{DM}$ colliders and direct detection experiments are likely to be complementary probes, sensitive to different parameter regions due to a different scaling behavior with the portal coupling $c_\phi$.
\item  If the $Z_2$ symmetry is approximate and only stabilizes $\phi$ on the timescale $\tau \gtrsim 10^{-8}$s but is allowed to decay in the early Universe, or if the $Z_2$ symmetry is exact but $\phi$ has hidden sector decays to other neutral states then colliders are the only probes of the Higgs Portal coupling, above or below threshold.
\end{itemize}

The Higgs Portal ties together aspects of cosmology and collider physics, allowing for very different probes of this coupling depending on the symmetry structure, mass, and thermal history of the Universe.  This interplay strongly motivates exploring the Higgs Portal over the widest mass range achievable. 

\subsection{Neutral naturalness}

Electroweak naturalness provides another motivation for the existence of neutral weak-scale scalars with large Higgs Portal couplings. Although most solutions to the hierarchy problem involve new states charged under the Standard Model, it is entirely possible that the weak scale is protected by additional degrees of freedom that are {\it neutral} under the Standard Model and couple exclusively through the Higgs Portal. Such states arise in the mirror twin Higgs \cite{Chacko:2005pe} and orbifold Higgs \cite{Craig:2014aea} models, and more generally are consistent with a bottom-up approach to naturalness \cite{Craig:2013xia}. 

A concrete, UV-complete realization of a Higgs Portal scenario relating to electroweak naturalness arises in the supersymmetric completion of the twin Higgs \cite{Chacko:2005pe, Chang:2006ra, Craig:2013fga}. Here the weak scale enjoys double protection from the approximate global symmetry of the twin Higgs mechanism as well as spontaneously broken supersymmetry. The role of the top partner is shared among the conventional supersymmetric partners of the top quark (the $\tilde t_L, \tilde t_R$), the SM-neutral fermionic top partners of the twin Higgs (the $t', \bar t'$), and the scalar superpartners of the $t'$s (the $\tilde t_L', \tilde t_R'$). Both the $t'$ and the $\tilde t'$ are pure singlets under the Standard Model and couple uniquely through the Higgs Portal. In particular, the scalars $\tilde t'$ inherit a coupling to the physical SM-like Higgs $h$ precisely of the form
\be
\mathcal{L} \supset |y_t|^2 v h (|\tilde t_L'|^2 + |\tilde t_R'|^2)  + \frac{1}{2} |y_t|^2 h^2 (|\tilde t_L'|^2 + |\tilde t_R'|^2) + \mathcal{O}(v^2/f^2)
\ee
where $f \gg v$ is the order parameter of global symmetry breaking in the twin Higgs. Here the sign of the coupling corresponds to double protection; the $\tilde t'$ serve to compensate for radiative corrections coming from the $\tilde t$ and $t'$. The $\tilde t_{L,R}'$ comprise six complex scalars in total, each with $\mathcal{O}(y_t^2)$ Higgs Portal coupling. If these states are approximately degenerate, then from the perspective of collider phenomenology this is equivalent to one real scalar with $|c_\phi| = \sqrt{3} |y_t|^2 \sim 1.7$. Although the detailed naturalness of this scenario depends on the mass scales of the $\tilde t, t'$, and $\tilde t'$, in general naturalness favors the $\tilde t'$ as close to the weak scale as possible.

\section{Searching for the Higgs Portal at $pp$ colliders} \label{sec:search}

Having motivated the parameter space for Higgs Portal interactions in a variety of scenarios, we now turn to $pp$-collider studies for the Higgs Portal model (\ref{eq:portal}) in various channels of interest at $\sqrt{s} = $ 14 \& 100 TeV. In this section we describe our collider simulation for searches involving vector boson fusion, monojet, and $t \bar t$ associated production, reserving a discussion of the results for Section \ref{sec:discussion}.
\par
For the signal events, we implement the model in \texttt{FeynRules}, setting $m_h = 125$ GeV. Events are then generated at leading order using \texttt{MadGraph5} \texttt{v1.5.8}~\cite{Alwall:2011uj}, fixing $c_\phi = 1$ and varying values of $m_\phi$. We infer results for $c_\phi \neq 1$ subsequently, by rescaling the signal cross section by $|c_\phi|^2$. We also simulate the primary backgrounds in \texttt{MadGraph5}. For both signal and backgrounds, the events are showered and hadronized using \texttt{Pythia\,8.186}~\cite{Sjostrand:2007gs}, tune 4C. Detector simulation is performed using \texttt{Delphes\,v3.1.2} with the default CMS detector card (for 14 TeV) and the Snowmass detector card ~\cite{Anderson:2013kxz} (for 100 TeV). Jets are clustered using the anti-k$_{\rm \tiny{T}}$ algorithm~\cite{Cacciari:2008gp}, as implemented in \texttt{FastJet\,v3.0.6}~\cite{Cacciari:2011ma}, with a cone size of $R=0.5$. All jets are required to have $p_{Tj}>$ 30 GeV. The lepton isolation criterion in $\texttt{Delphes}$ is defined as $\mathtt{RelIso} \equiv p_T^{\rm cone} / p_{T \ell} < 0.1$, where $p_T^{\rm cone}$ is the sum of hadronic $p_T$ within a cone of $R= 0.3$ of the lepton. A minimal $p_T$ cut of $10$ GeV is applied for all leptons. 

\subsection{The Higgs Portal in $\met \,+ \,$vector boson fusion} \label{sec:vbf}

We begin with vector boson fusion, which we expect will be the primary discovery channel for scalars coupling through the Higgs Portal. The topology for this process is identical to that of an invisibly-decaying Higgs produced via vector boson fusion, save that now the intermediate Higgs is pushed off-shell. The final state is $\phi \phi j j$ with forward jets, while the primary backgrounds to this process are $Z$+jets, $W$+jets, $t \bar t$ + jets, and QCD multijets. For this search we simulate $Zjj$ and $Wjj$ matched up to one additional jet and $t \bar t$ matched up to two additional jets. We do not simulate QCD multijets due to the usual challenges of reliably simulating multijet production, but we adopt a cut flow designed to minimize QCD multijet backgrounds.
\par
After requiring at least two jets in the event, we apply the following baseline cuts
\begin{eqnarray}
&& p_{Tj_{1(2)}} > 50 \, {\rm GeV} \hspace{1cm} |\eta_{j_{1(2)}}| < 4.7 \\
&& \eta_{j_1} \eta_{j_2} < 0 \hspace{1cm} |\eta_{j_1} - \eta_{j_2}| > 4.2
\end{eqnarray}
In addition to these cuts, we veto events containing an isolated $e^\pm$ or $\mu^\pm$, using the isolation requirement as defined earlier. We also apply a central-jet veto by vetoing events containing a third jet with $p_{Tj} > 30$ GeV and ${\rm min}\,\eta_{j_{1,2}} < \eta_{j3} < {\rm max}\,\eta_{j_{1,2}}$. To isolate the signal, we apply both a dijet invariant mass cut and a $\met$ cut:
\begin{eqnarray}
\sqrt{(p_{j_1}+p_{j_1})^2} > M_{jj}^* \hspace{1cm} \met > \met^*
\end{eqnarray}
Here, $M_{jj}^*$ and $\met^*$ are partially optimized values for the dijet invariant mass and $\met$ cuts, chosen at each value of $m_\phi$, so as to maximize $S/\sqrt{B}$. Finally, a cut on the azimuthal angle between $\met$ and jets is imposed by demanding $|\Delta \phi_{\met, j}| > 0.5$. This cut has negligible effects on the results in our case, but is included to ensure that QCD backgrounds are sufficiently suppressed in realistic scenarios.
\par
For $\sqrt{s} = 14$ TeV we attempt to account for the anticipated effects of pileup. We simulate pile-up events by overlaying $N_a$ soft-QCD events, drawn from a Poisson distribution with mean $\langle N \rangle_{{\rm \tiny{PU}}} = 100$, for each event $a$. The soft-QCD events are generated in $\mathtt{Pythia\,8.186}$~\cite{Sjostrand:2007gs}. We find that the inclusion of pileup in this manner roughly decreases the significance by a factor of $2-3$ across different values of $m_\phi$. Given that the expected pileup and performance of jet-grooming algorithms is entirely unknown for future colliders, we do not estimate the effects of pileup at $\sqrt{s} = 100$ TeV.

\subsection{The Higgs Portal in $\met +j$ associated production}\label{sec:ggf}

\begin{figure}[tbp] 
   \centering
 \includegraphics[width=2.0in]{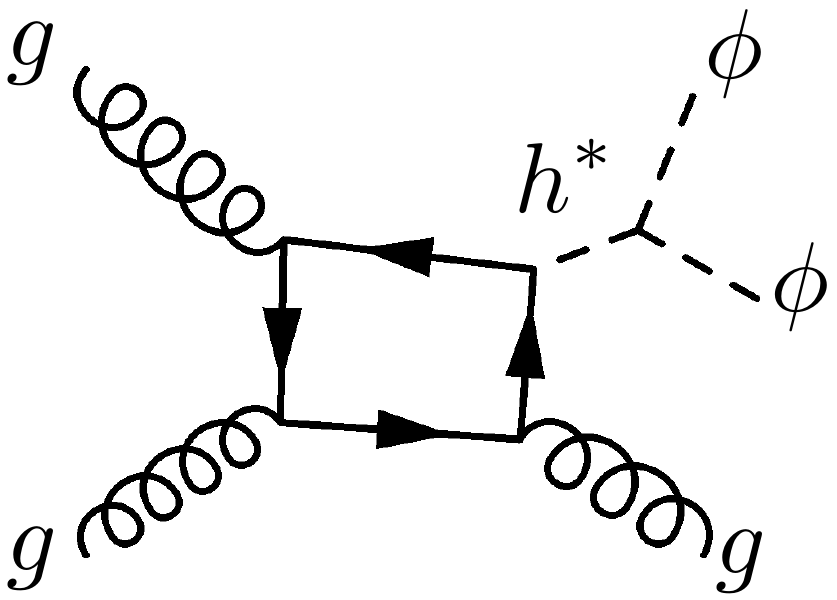}
 {  \caption{An example of the loop processes contributing to the $\met +j$ signal from gluon associated production at hadron colliders.}
   \label{fig:loop}}
\end{figure}

Next, we consider the sensitivity of searches for the Higgs Portal in the $j+\met$ channel via gluon fusion with an associated jet. A sample diagram for this channel is depicted in \Fig{fig:loop}. Although this channel sets a sub-leading limit at $\sqrt{s} = 8$ TeV \cite{Endo:2014cca}, the increasing gluon partonic luminosity at higher center-of-mass energies makes it a promising channel for future colliders. The primary backgrounds for this process are again $Z$+jets, $W$+jets, $t \bar t$ + jets, and QCD multijets. Here we simulate $Zj$ and $Wj$ matched up to one additional jet and $t \bar t$ matched up to two additional jets, and again do not simulate QCD multijets but adopt a cut flow designed to minimize this background.

\begin{figure}[htbp] 
   \centering
 \includegraphics[width=3in]{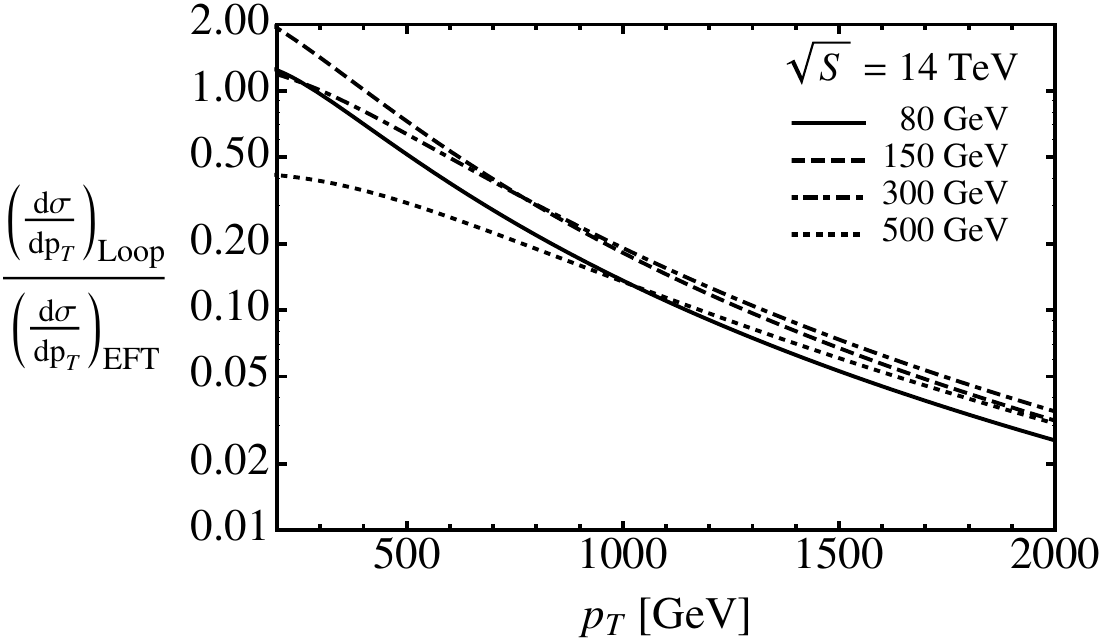} 
 \includegraphics[width=3in]{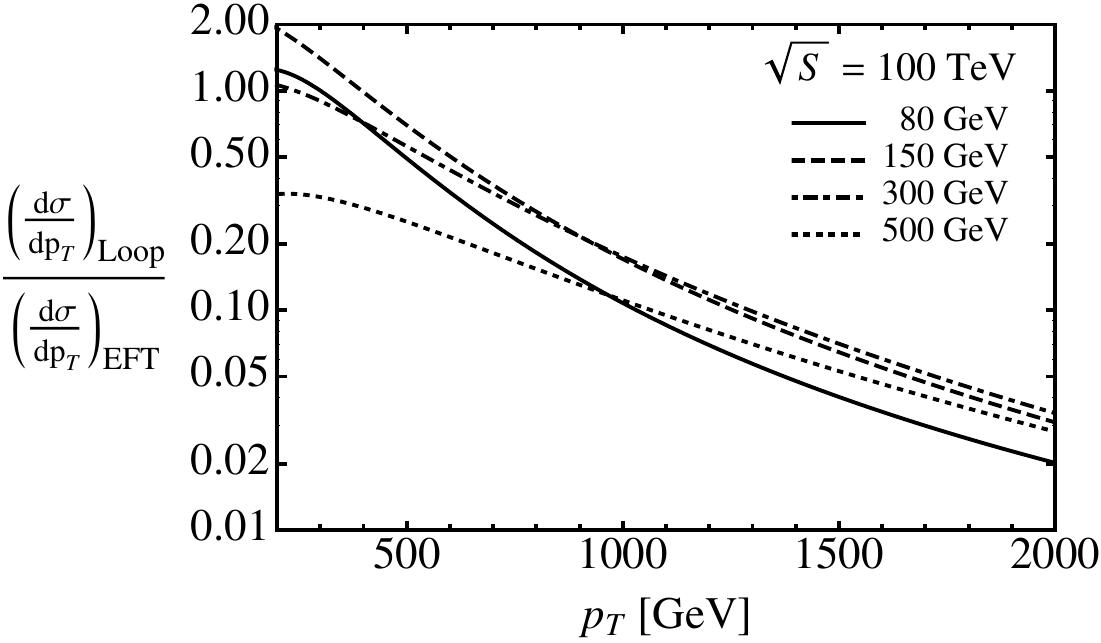}
 {  \caption{The ratio of the differential cross section for $j+\met$ for the full one-loop result relative to the HEFT result at 14 and 100 TeV for a variety of singlet scalar masses.  When $p_T \sim 2 m_t$ there can be an $\mathcal{O}(1)$ enhancement due to the top mass threshold in the loop.  At higher $p_T > 2 m_T$ the HEFT calculation may overestimate the signal rate significantly.}
   \label{fig:lvseft}}
\end{figure}

As $\sqrt{s} \gg 2 m_t$ for the majority of signal events, the HEFT calculation of $gg\to g h^* \to g \phi \phi$ (which is accurate only to lowest order in $1/m_t^2$) is not valid.  To correct for this we perform a $p_T$-dependent reweighting of signal events generated using the HEFT in \texttt{MadGraph5}.\footnote{For other recent approaches to this problem, see \cite{Buckley:2014fba,Harris:2014hga}.}  For the reweighting factor the differential cross section for $gg\to g \phi \phi$ was calculated from the cross section for $gg\to g h^*$ using the factorization of phase space due to the scalar Higgs propagator
\be
\frac{d \sigma^{L,EFT}_{gg\to g \phi \phi} (m_\phi)}{d p_T} = \int^\infty_{4 m_\phi^2} \frac{c_\phi^2}{8 \pi^2}  \frac{v^2}{(\tilde{s}-m_h^2)^2}  \sqrt{1-\frac{4 m_\phi^2}{\tilde{s}}}  \frac{d \sigma^{L,EFT}_{gg\to g h^*}}{d p_T} d \sqrt{\tilde{s}} ~~,
\ee
where the superscripts `L' and `EFT' denote the one loop and EFT cross sections and $v=246$ GeV.  These parton-level cross sections were convoluted with the MSTW pdfs \cite{Martin:2009iq} to determine the proton-proton differential cross section.  In a given $p_T$-bin the reweighting factor is defined as
\be 
R(p_T^{\rm min},p_T^{\rm max},m_\phi)=\frac{\int^{p_T^{\rm max}}_{p_T^{\rm min}} d p_T \frac{d \sigma^{L}_{pp\to g \phi \phi}  (m_\phi)}{d p_T}}{\int^{p_T^{\rm max}}_{p_T^{\rm min}} d p_T \frac{d \sigma^{EFT}_{pp\to g \phi \phi}  (m_\phi)}{d p_T}}
\label{eq:RWfact}
\ee
The EFT cross section $\sigma^{EFT}_{gg\to g h^*}$ was calculated using the results of \cite{Ellis:1987xu,Dawson:1990zj} and the cross section incorporating the full loop functions, $\sigma^{L}_{gg\to g h^*}$, was calculated using the {\sc{FeynArts}}, {\sc{FormCalc}}, and {\sc{LoopTools}} suite of packages~\cite{Hahn:2000kx,Hahn:1998yk}.  The renormalization and factorization scales were set to the partonic CM energy.  In the limit of small partonic CM energy it was checked that the partonic EFT and loop calculations match as expected.  As demonstrated in \Fig{fig:lvseft}, for high CM energies the EFT may overestimate the cross section significantly, thus the suppression factor is significant.  Also, when processes at $\sqrt{s} \sim 2 m_t$ contribute significantly to the signal phase space the EFT calculation may underestimate the signal by $\mathcal{O}(10\text{'s} \%)$. 
\par
To investigate the search sensitivity in this channel, we require at least one jet in the event and apply the following cuts to signal and background at $\sqrt{s} = 14, 100$ TeV:
\begin{equation}
p_{Tj_1} > 110 \, {\rm GeV} \hspace{1cm} |\eta_{j_1}| < 2.4 \hspace{1cm} \met > 300 \, {\rm GeV}
\end{equation}
The restrictive $\eta_{j_1}$ cut is chosen pragmatically to expedite the calculation of re-weighting factors, and in practice could be relaxed. Since we do not have enough computational power for generating enough QCD background events, we include an additional jet veto analogous to the ones applied in CMS monojet searches \cite{Khachatryan:2014rra}. A second jet with $p_{Tj_2} > 30$ GeV is allowed as long as $\Delta R_{j_1, j_2} <2.4$, otherwise the event is vetoed. Events containing additional jets with $p_{Tj} > 30$ GeV are vetoed, as are events containing an isolated lepton candidate. It is possible that QCD multi-jet backgrounds at $\sqrt{s} = 100$ TeV will favor harder $\met$ cuts than those applied here, but reliable simulation of such backgrounds is beyond the capacity of this study.
\par
To compensate for inadequacies in the HEFT approximation in the event generation, we finally apply the appropriate re-weighting factors as defined in (\ref{eq:RWfact}). Due to the relatively rapid fall-off of the jet $p_T$ spectrum in the re-weighted signal events, no meaningful improvement of signal significance can be obtained from applying harder $p_T$ and $\met$ cuts to the simulated backgrounds as a function of $m_\phi$. 

\subsection{The Higgs Portal in $\met +t \bar t$ associated production} \label{sec:tth}

Finally, we consider the sensitivity of searches for the Higgs Portal in the $t \bar t + \met$ channel. This channel sets a promising limit on invisible Higgs decays at $\sqrt{s} = 8$ TeV \cite{Zhou:2014dba}, suggesting it may potentially be interesting in future Higgs Portal searches at the LHC and beyond. The dominant backgrounds in this channel are expected to be $t\bar{t}$+jets and $W$+jets. To improve statistics, we separately simulate semi-leptonic and di-leptonic decays for the $t \bar t$ background matched up to two additional jets, while we simulate leptonic $Wjj$ matched up to two additional jets. To extract the sensitivity in this channel, we first apply the following requirements:
\begin{eqnarray}
&& n_{\rm jet} \ge 4 \hspace{1cm}
|\eta_{j_{1,2,3,4}}| < 2.4 \hspace{1cm}
\met > 300 \,\, {\rm GeV}
\end{eqnarray}
In addition, we require exactly one isolated $e^\pm/\mu^\pm$ with $$P_T^{\ell} > \,10\,\,\rm{GeV}$$ and at least one $b$-tag among the leading four jets. We  also require that the transverse mass between the lepton and $\met$ is constrained to $m_T > 200$ GeV and that $M_{T2}^W>200\,\rm{GeV}$~\cite{Bai:2012gs}. 

\section{Results and Discussion} \label{sec:discussion}

\begin{figure}[htbp] 
   \centering
 \includegraphics[width=3in]{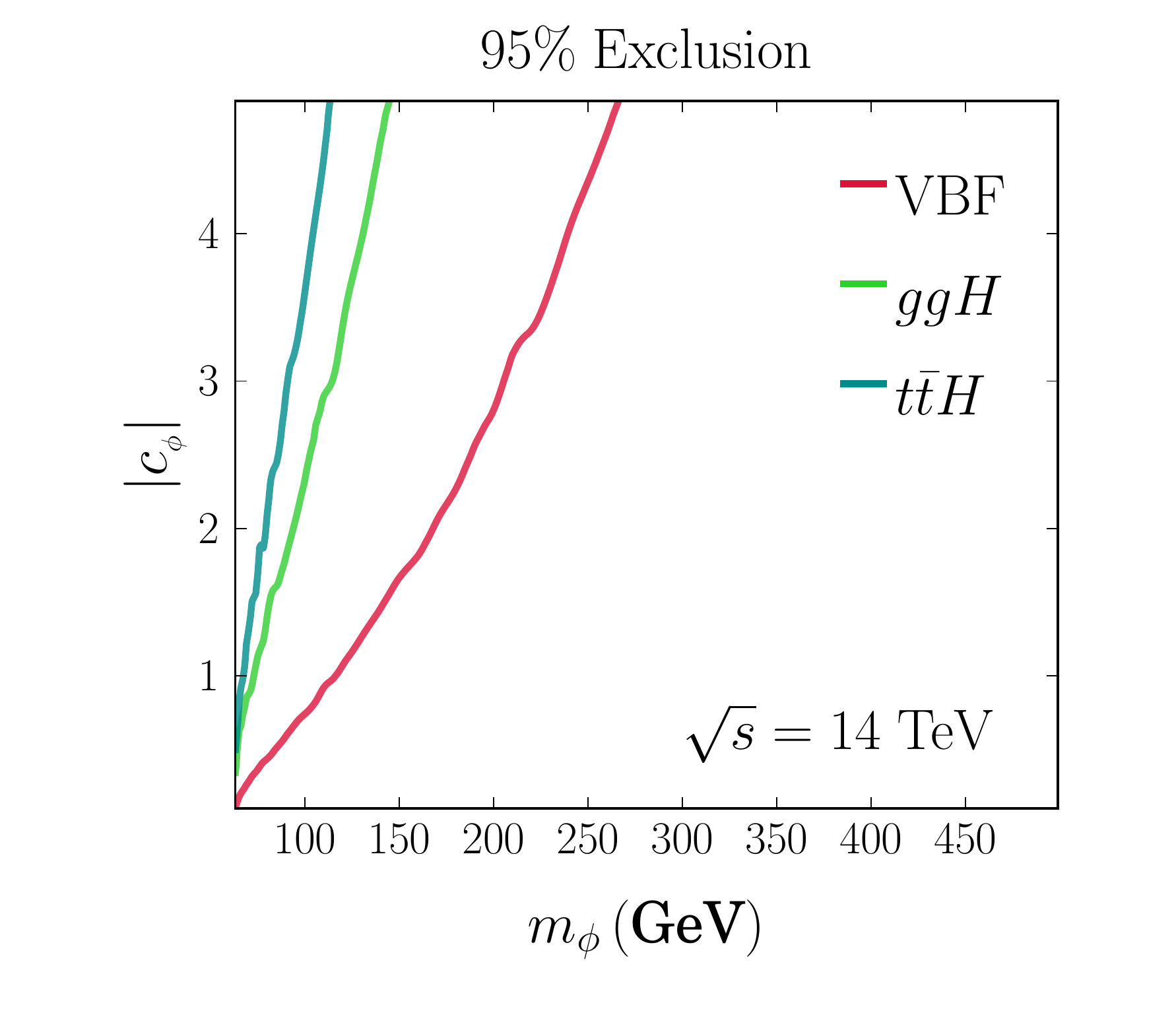} 
 \includegraphics[width=3in]{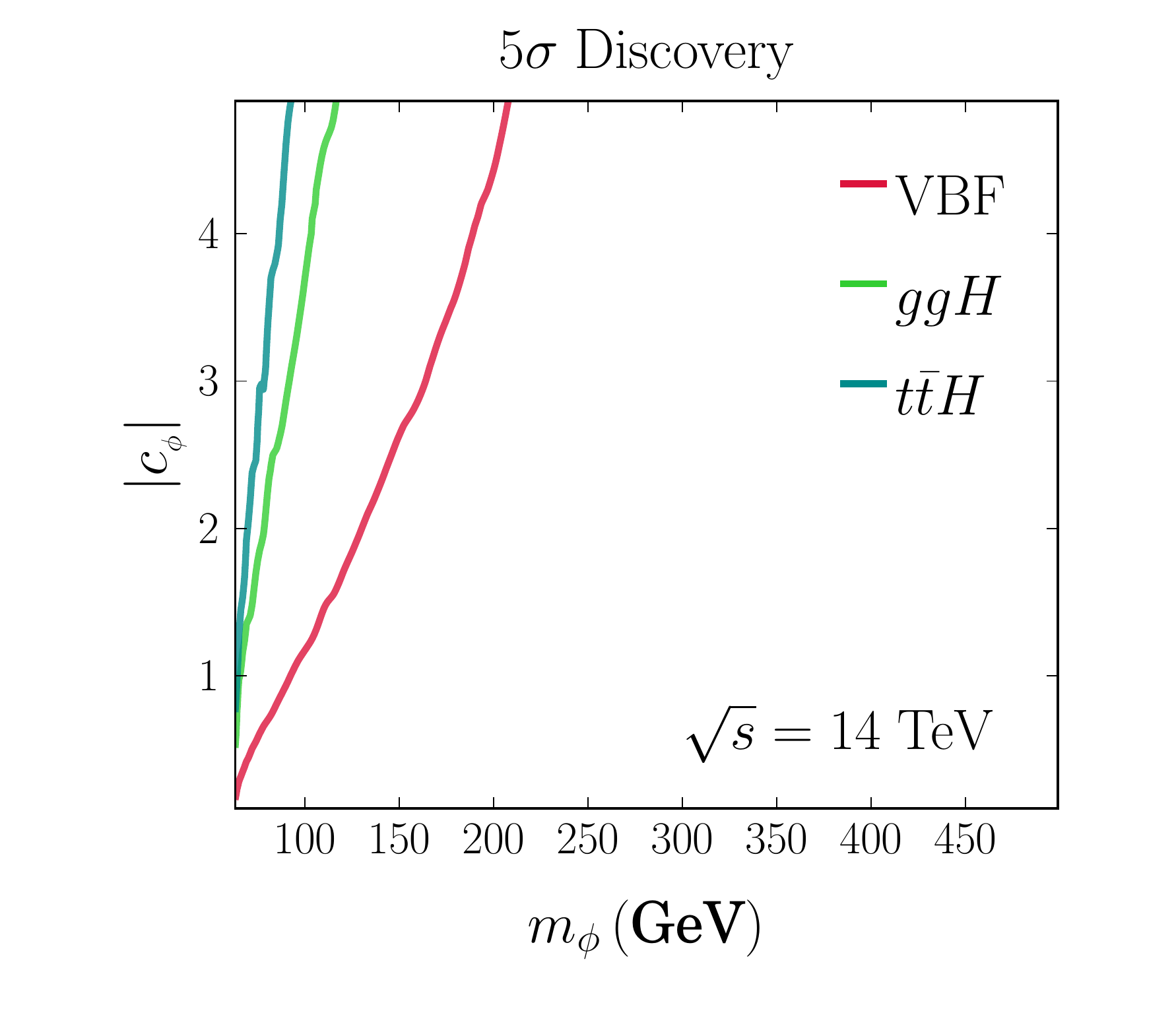}
{   \caption{Left: 95$\%$ exclusion reach in all three channels at $\sqrt{s} = 14$ TeV determined from $S/\sqrt{S+B} = 1.96$, neglecting systematic errors.  Right: 5$\sigma$ discovery reach in the VBF and monojet channels at $\sqrt{s} = 14$ TeV determined from $S/\sqrt{B} = 5$, again neglecting systematic errors. }
   \label{fig:14}}
\end{figure}

\begin{figure}[htbp] 
   \centering
 \includegraphics[width=3in]{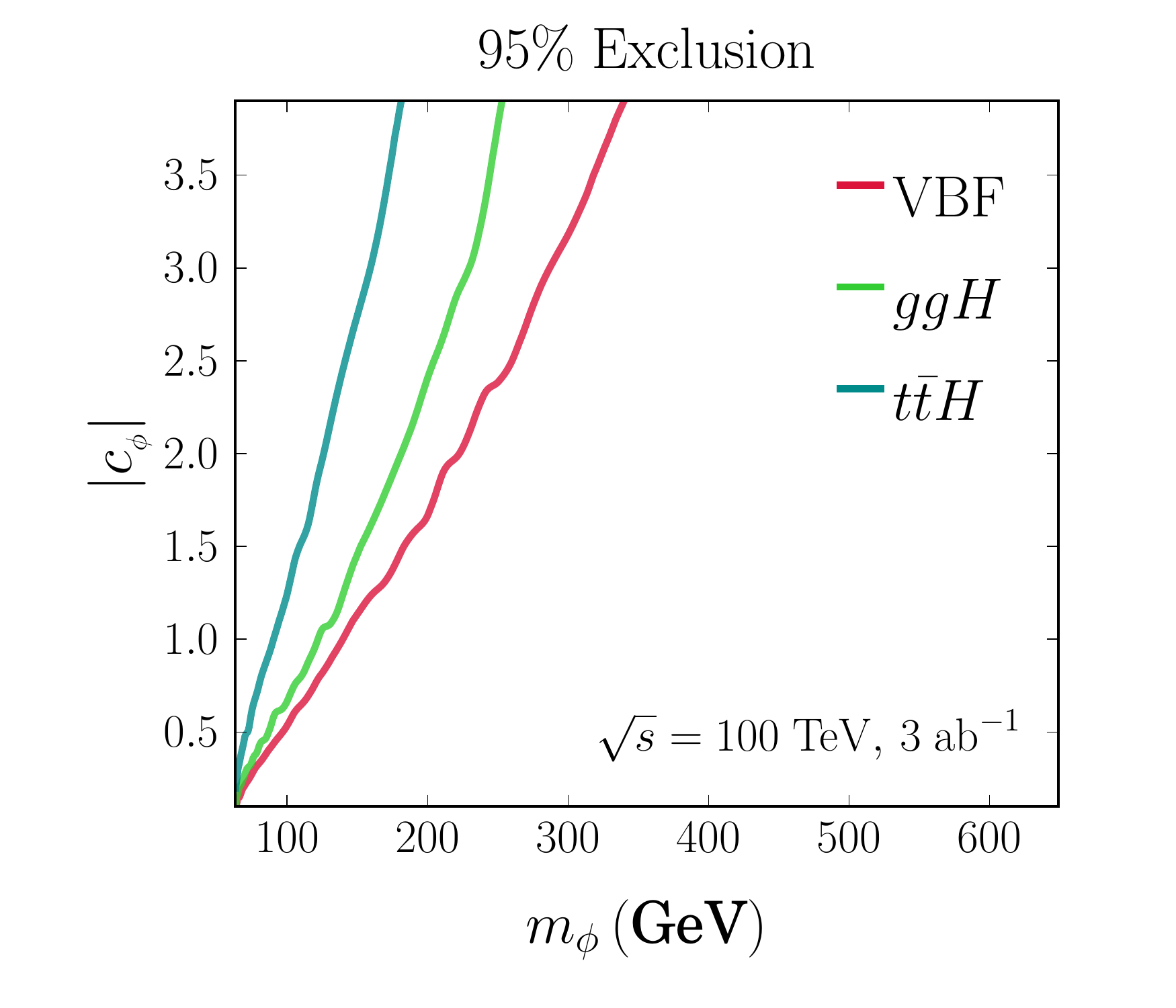} 
 \includegraphics[width=3in]{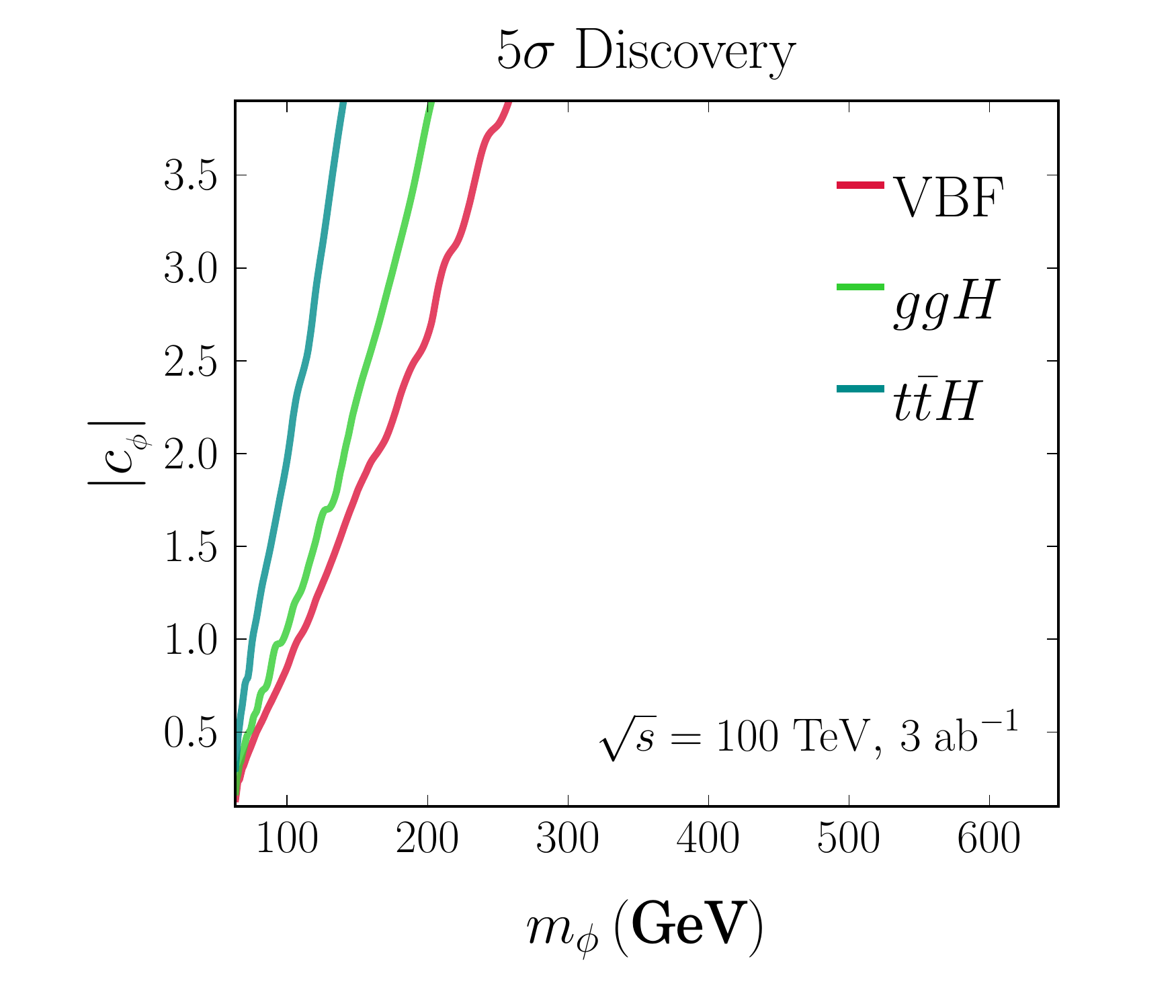}
{    \caption{Left: 95$\%$ exclusion reach in all three channels with 3 ab$^{-1}$ at $\sqrt{s} = 100$ TeV determined from $S/\sqrt{S+B} = 1.96$, neglecting systematic errors.  Right: 5$\sigma$ discovery reach in the VBF and monojet channels with 3 ab$^{-1}$ at $\sqrt{s} = 100$ TeV determined from $S/\sqrt{B} = 5$, again neglecting systematic errors.  }
 \label{fig:100ab3}}
\end{figure}

\begin{figure}[htbp] 
   \centering
 \includegraphics[width=3in]{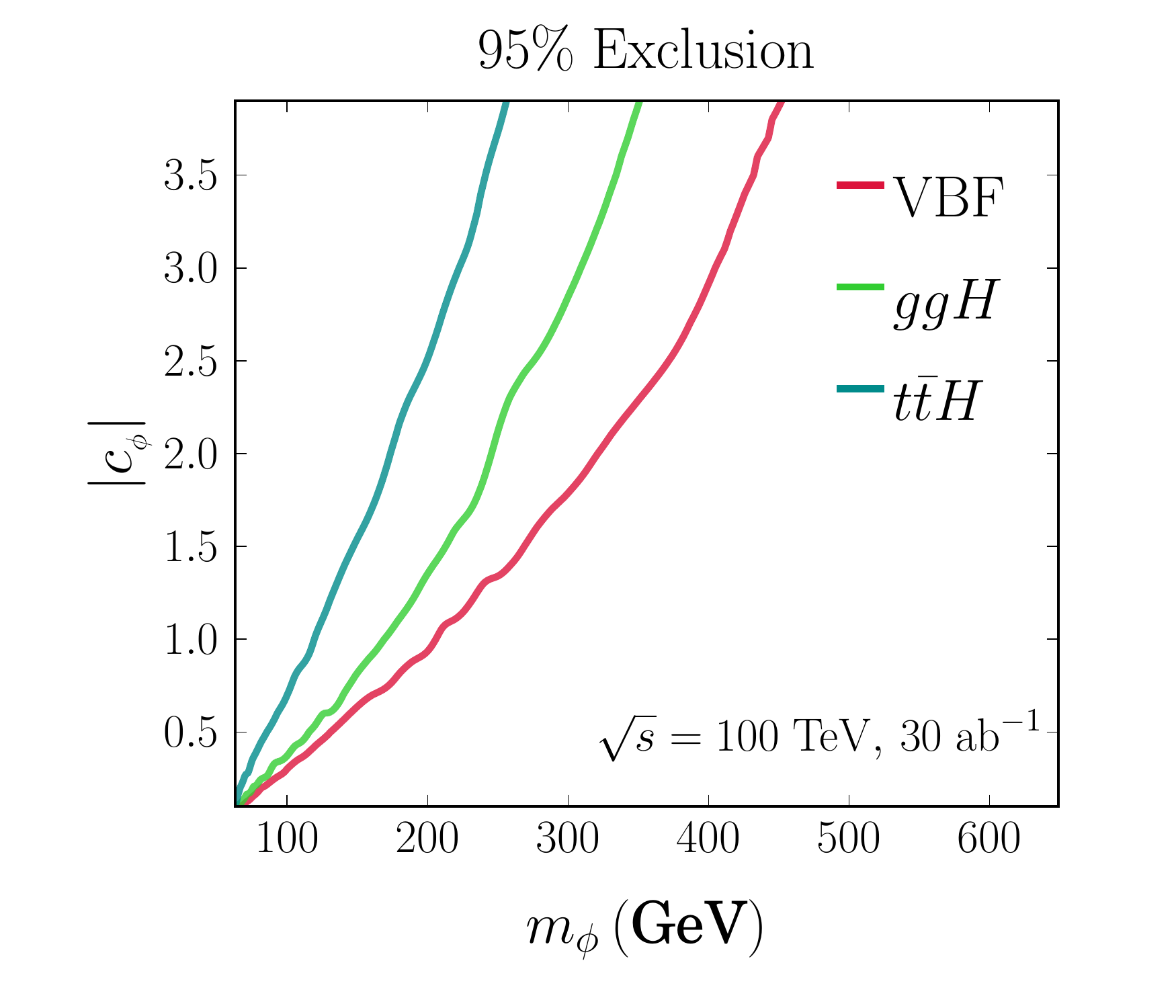} 
 \includegraphics[width=3in]{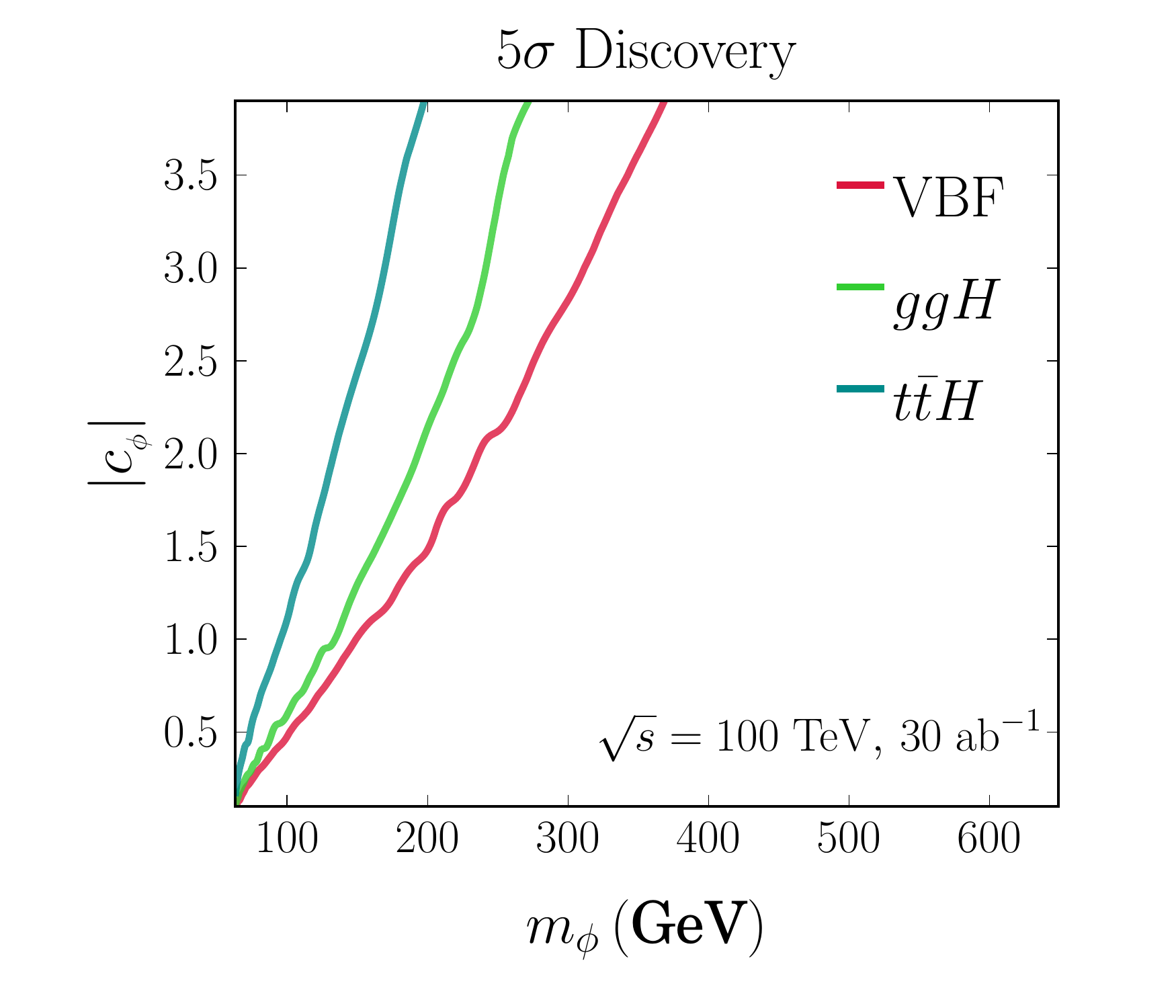}
{    \caption{Left: 95$\%$ exclusion reach in all three channels with 30 ab$^{-1}$ at $\sqrt{s} = 100$ TeV determined from $S/\sqrt{S+B} = 1.96$, neglecting systematic errors.  Right: 5$\sigma$ discovery reach in the VBF and monojet channels with 30 ab$^{-1}$ at $\sqrt{s} = 100$ TeV determined from $S/\sqrt{B} = 5$, again neglecting systematic errors.  }
 \label{fig:100ab30}}
\end{figure}

\begin{figure}[htbp] 
   \centering
 \includegraphics[width=3in]{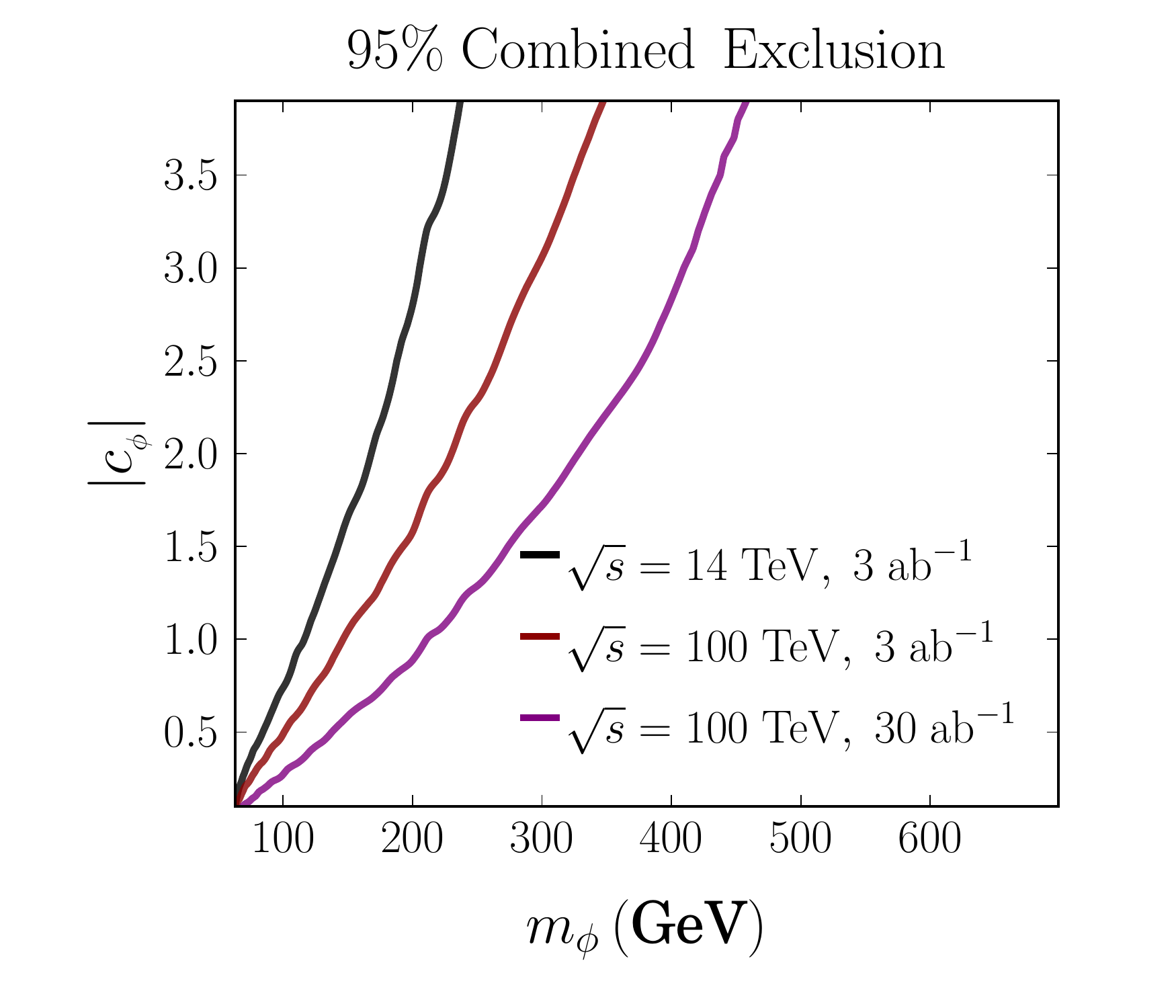} 
  \includegraphics[width=3in]{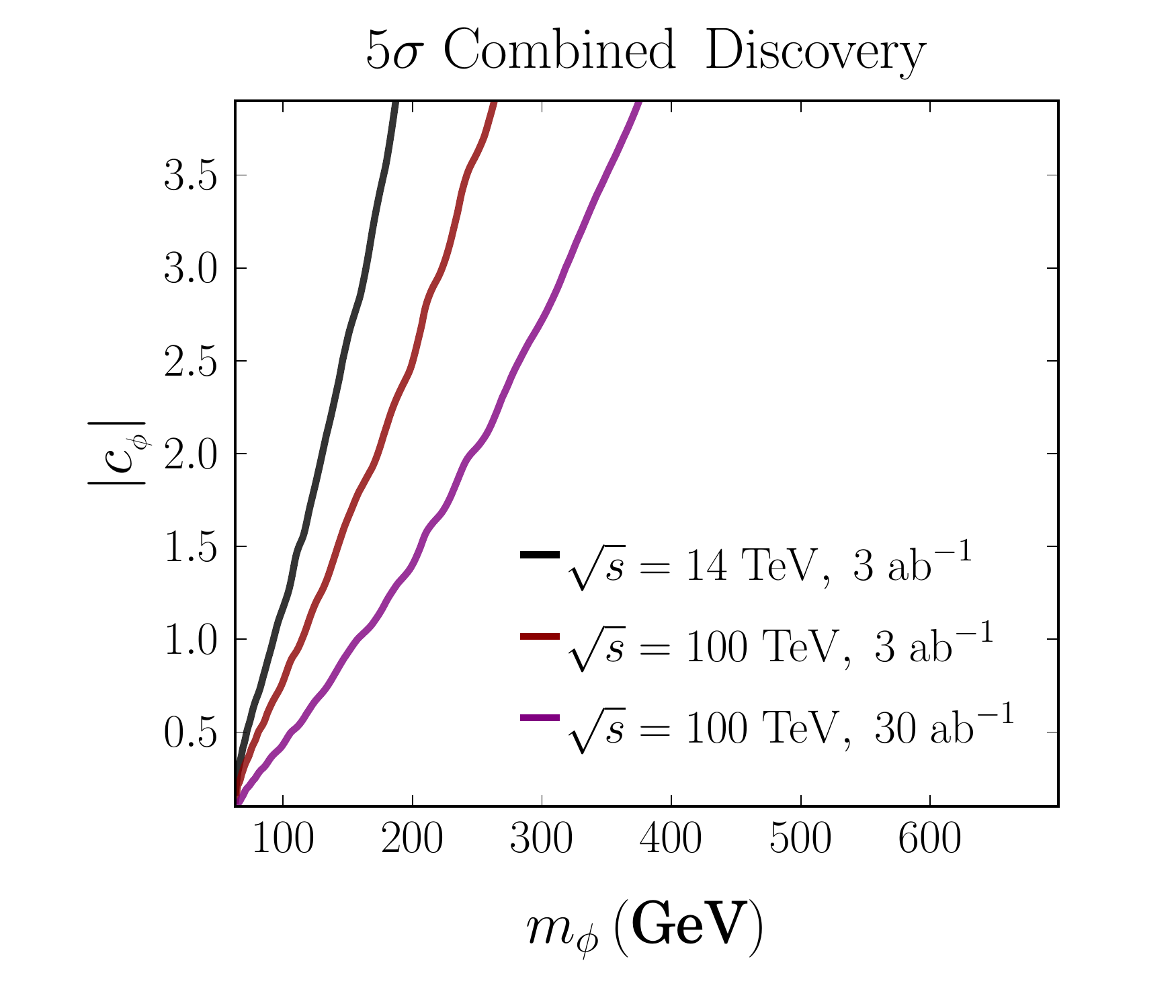} 
{   \caption{Left: Approximate 95$\%$ exclusion reach from the combination of VBF, $ggH$ and $t\bar{t}H$ channels with 3 ab$^{-1}$ at $\sqrt{s} = 14$ and 3, 30 ab$^{-1}$ at $\sqrt{s} = 100$ TeV determined from $S/\sqrt{B} = 1.96$, neglecting systematic errors and correlations between channels. Right: Approximate 5$\sigma$ discovery reach from the same combination at $\sqrt{s} = 14, 100$ TeV.}
   \label{fig:combined}}
\end{figure}

We have performed a simple cut and count analysis following the cut flows for the searches outlined in Sections \ref{sec:vbf}, \ref{sec:ggf}, and \ref{sec:tth}. For $\sqrt{s} = $ 14 TeV we assume an integrated luminosity of 3 ab$^{-1}$. For $\sqrt{s} =$ 100 TeV we consider scenarios with 3 ab$^{-1}$ and 30 ab$^{-1}$, respectively. We compute the significance of a search in terms of signal events $S$ and background events $B$ passing cuts as
\begin{equation}
{\rm Exclusion \;\,Sign.} = \frac{S}{\sqrt{S+B }}
\hspace{1cm}
{\rm Discovery \;\,Sign.} = \frac{S}{\sqrt{B }}
\end{equation}
neglecting systematic uncertainties in the signal and background estimates.  In principle, systematic uncertainties in background determination could have a substantial impact at $\sqrt{s} = 100$ TeV since $S/B$ is quite small. However, in practice one expects data-driven determination of $Z$+jets and other backgrounds to substantially lower systematic uncertainties by the 100 TeV era.

Results for the exclusion and discovery reach of the VBF, monojet, and $t \bar t$ searches at $\sqrt{s} = 14$ TeV are presented in Fig.~\ref{fig:14}. For the VBF channel at $14$ TeV, our preliminary study of pileup effects indicates that $S/\sqrt{B}$ is reduced approximately by a factor of $2-3$ for $\langle N \rangle_{\text{\tiny{PU}}}\sim 100$. This may potentially be mitigated through the use of next-generation jet-grooming algorithms~(see for example ~\cite{Cacciari:2007fd,Krohn:2013lba,Berta:2014eza}). As expected, all three channels improve significantly over the $\sqrt{s} = 8$ TeV VBF reach, while the VBF channel substantially outperforms the monojet and $t \bar t$ channels at $\sqrt{s} = 14$ TeV.

The corresponding results for VBF, monojet, and $t \bar t$ searches at $\sqrt{s} = 100$ TeV are presented in Figs.~\ref{fig:100ab3} and \ref{fig:100ab30} for the 3 ab$^{-1}$ and 30 ab$^{-1}$ scenarios, respectively. Here we do not include pileup estimates, as the operating parameters and efficacy of jet-grooming algorithms are entirely unknown. The reach of the VBF search is in fairly good agreement with the simplified analysis in \cite{Curtin:2014jma}, with a modest reduction in sensitivity due to the additional backgrounds considered here. Surprisingly, at $\sqrt{s} = 100$ TeV the monojet search exhibits more comparable sensitivity to the VBF search for $m_\phi \lesssim$ 200 GeV, due in part to the effects of increased gluon PDF luminosity and the relatively low jet $p_T$ cuts relative to the center of mass energy. We caution that some of this sensitivity may be an artifact of our fixed-order calculation for the monojet signal, and furthermore neglects possible contributions from QCD multijet backgrounds that may be appreciable in this case. On the other hand, we have not included a $K$-factor for gluon fusion associated production, which would further enhance sensitivity. On the whole, our results suggest that the monojet + missing energy channel may be useful at $\sqrt{s} = 100$ TeV and warrants further study. In contrast, the $t \bar t$ associated production search demonstrates relatively poor sensitivity, though there is substantial room for improvement through the use of more sophisticated discriminating variables such as hadronic chi-square \cite{Chatrchyan:2013xna}.

To estimate the reach of a concerted Higgs Portal search program, we present the approximate combined reach of VBF, monojet, and $t \bar t$ searches at $\sqrt{s} = 14$ and 100 TeV  in  Fig.~\ref{fig:combined}. We obtain the combination by adding the significance of the VBF,  monojet, and $t \bar t$ channels in quadrature, neglecting possible correlations between the two channels. 

\subsection{Implications for new physics}

\begin{figure}[htbp] 
   \centering
 \includegraphics[width=3in]{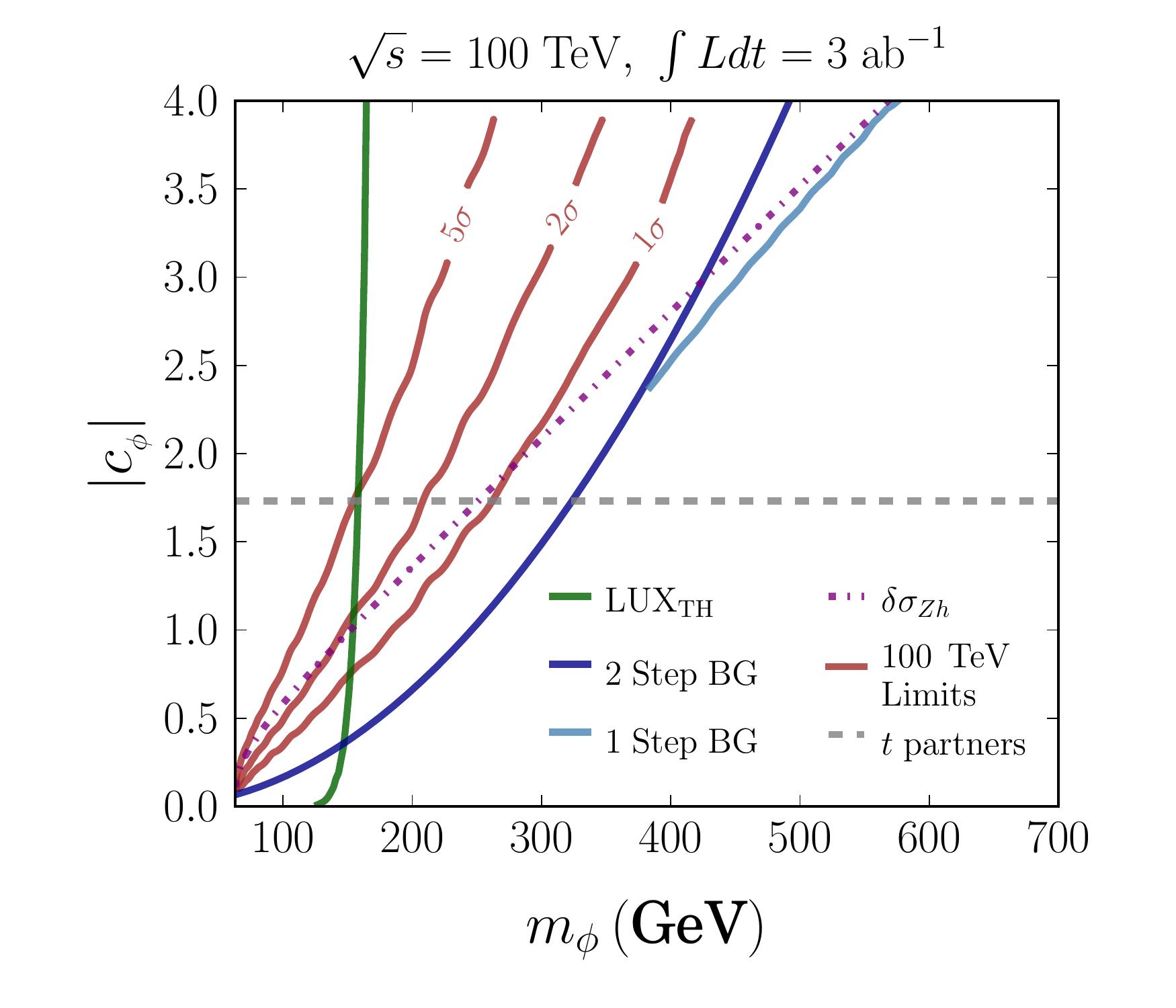}
  \includegraphics[width=3in]{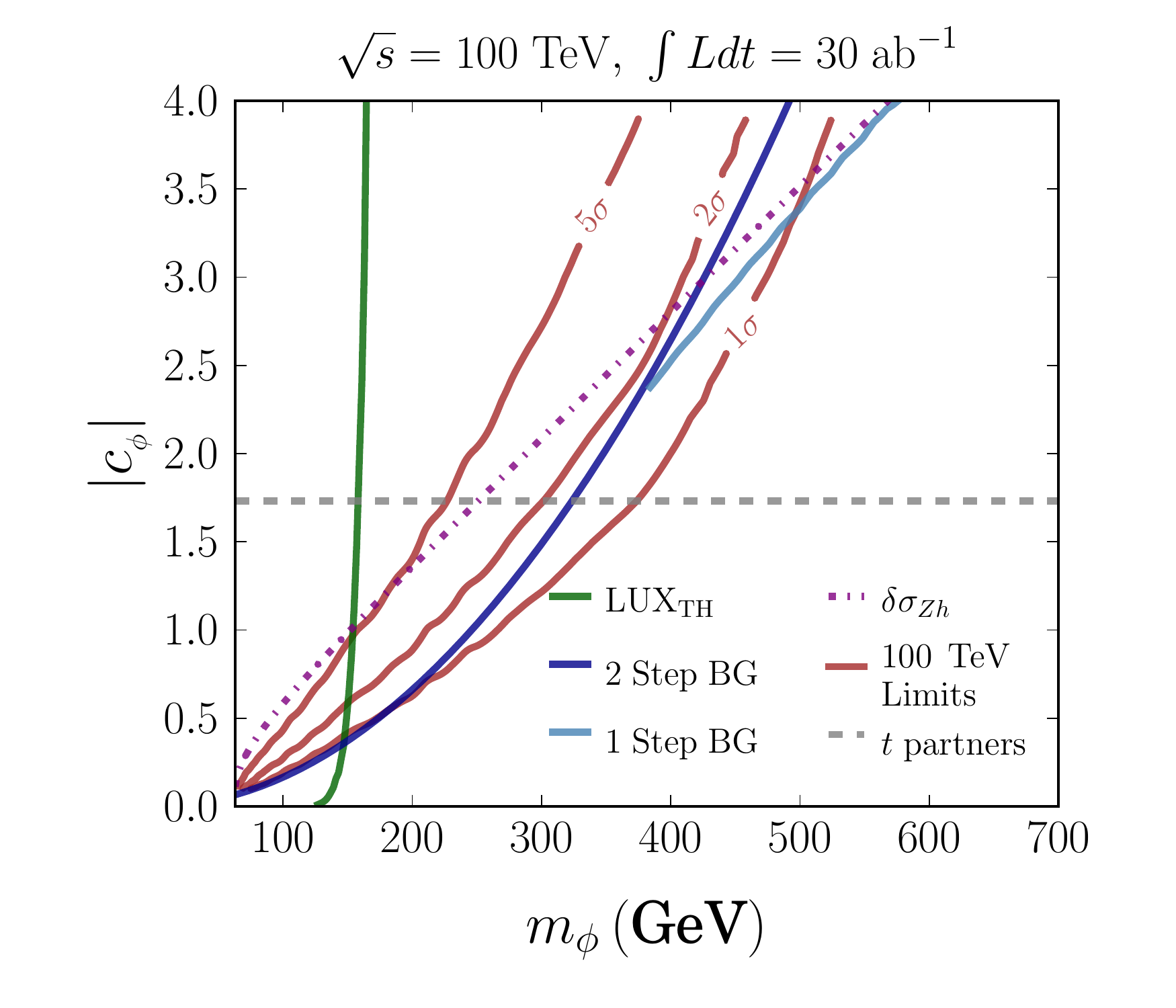}  
{   \caption{Combined reach of direct searches in VBF, $ggH$ and $t\bar{t}H$ channels at $\sqrt{s} = 100$ TeV for 3 ab$^{-1}$ (left) and 30 ab$^{-1}$ (right) compared to select parameter spaces for motivated Higgs Portal scenarios. In each plot the red lines denotes the $1\sigma$ exclusion, $2\sigma$ exclusion, and $5\sigma$ discovery reach from direct searches at $\sqrt{s} = 100$ TeV. The region to the left of the green line denotes the LUX exclusion for Higgs Portal dark matter with thermal abundance given by $c_\phi, m_\phi$. The region to the left of the dark blue line denotes the possible parameter space for two-step baryogenesis, while the region between the light blue and dark blue lines denotes the possible parameter space for one-step baryogenesis (defined by $v_c /T_c \geq 0.6$). The purple line denotes the 2$\sigma$ contour for $\delta \sigma_{Zh}$ at a future $e^+e^-$ circular collider such as TLEP.  The dashed gray line denotes the effective coupling of six complex scalar top partners.}
   \label{fig:summary}}
\end{figure}

Although any reach in the Higgs Portal parameter space is valuable, it is useful to compare the exclusion and discovery reach of searches for Higgs Portal states at $pp$ colliders to the range of masses and couplings motivated by the BSM scenarios discussed in Section \ref{sec:np}. In Fig.~\ref{fig:summary} the combined reach at 100 TeV for both 3 and 30 ab$^{-1}$ is shown relative to both the reach of Higgs coupling measurements at a circular $e^+ e^-$ collider and the motivated parameter space for electroweak baryogenesis, dark matter, and neutral naturalness.

In the case of singlet-assisted electroweak baryogenesis, the combined $2 \sigma$ exclusion reach at $\sqrt{s} = 100$ TeV with 30 ab$^{-1}$ covers most of the region of the double-step phase transition. While we cannot decisively exclude some parts of the double-step phase transition or the single-step phase transition with our analysis, there is clearly some sensitivity throughout the region of viable EWBG as evidenced by the $1 \sigma$ exclusion contour. Given this sensitivity, it may well be that optimized searches at 100 TeV can conclusively exclude (or possibly discover) the scenario. To the extent that this represents the most observationally-challenging scenario for a strongly first-order phase transition, a 100 TeV collider may be well-positioned to decisively settle the question of electroweak baryogenesis.
 
In the case of dark matter, collider searches for Higgs Portal states are not competitive with dark matter direct detection for small couplings, but at $c_\phi \gtrsim 1$ can exceed the exclusion and discovery reach of the LUX direct detection experiment when the Higgs portal state possesses its natural thermal abundance. In the event of a signal in future direct detection experiments, this also suggests that direct evidence for Higgs Portal states may be obtained through searches at colliders.

In the case of neutral naturalness, the $2 \sigma$ exclusion reach extends out to neutral top partners with $m_\phi \sim 300$ GeV with 30 ab$^{-1}$ at $\sqrt{s} = 100$ TeV. This corresponds to a fine-tuning of the weak scale on the order of 30\% from the neutral top partners alone \cite{Craig:2014aea}, and in complete models with neutral top partners the fine-tuning is expected to be considerably worse. Considering that this scenario represents the worst-case scenario for electroweak naturalness (from the perspective of collider signatures), pushing naturalness to the 30\% level in this case represents an impressive achievement.

Finally, we can compare the combined reach at 100 TeV to the sensitivity of indirect probes of the Higgs Portal via shifts in the $Zh$ production cross section. The leading order shift due to (\ref{eq:portal}) is \cite{Englert:2013tya,Craig:2013xia}
\begin{equation}
\delta \sigma_{Zh} = \frac{ |c_\phi|^2}{8 \pi^2} \frac{v^2}{m_h^2} \left(1 + \frac{1}{4 \sqrt{\tau_\phi(\tau_\phi -1)}} \log \left[ \frac{1 - 2 \tau_\phi - 2 \sqrt{\tau_\phi (\tau_\phi-1)}}{1-2 \tau_\phi + 2 \sqrt{\tau_\phi(\tau_\phi - 1)}} \right]  \right)
\end{equation} 
where $\tau_\phi = m_h^2 / 4 m_\phi^2$ and $\delta \sigma_{Zh} = (\sigma_{Zh} - \sigma_{Zh}^{\rm SM})/  \sigma_{Zh}^{\rm SM}$. 
In Fig.~\ref{fig:summary} we compare the $2 \sigma$ reach at a future $e+ e-$ machine such as CEPC/TLEP to the $2 \sigma$ exclusion reach and $5 \sigma$ discovery reach at a 100 TeV $pp$ machine, with an eye towards determining whether observed deviations in the $Zh$ cross section may lead to the discovery of new singlet states. We use the Snowmass projections for TLEP sensitivity at $\sqrt{s} = 240$ GeV \cite{Dawson:2013bba}. The $2 \sigma$ exclusion reach of a 100 TeV machine is comparable to the equivalent reach at a circular $e^+ e^-$ throughout the parameter space under consideration, with direct searches performing better at small couplings and $Zh$ precision performing better at larger coupings. Compellingly, we find that the 5$\sigma$ discovery reach at a 100 TeV $pp$ machine with 30 ab$^{-1}$ also exceeds the $2 \sigma$ reach at a circular $e^+ e^-$ collider up to $m_\phi \sim 200$ GeV, making a 100 TeV $pp$ machine a powerful tool for direct discovery of a high-mass Higgs Portal in the event of suggestive hints in precision Higgs coupling measurements. Moreover, it implies that for $m_\phi \lesssim 200$ GeV, a 100 TeV machine is capable of discovering a high-mass Higgs Portal even in the absence of suggestive deviations in precision Higgs measurements.

\section{Conclusions} \label{sec:conclusions}

The discovery of the Higgs boson brings forth qualitatively new opportunities in the search for physics beyond the Standard Model. The Higgs Portal is one of the most salient such opportunities. While efforts so far have focused on the production of Higgs Portal states through the decay of an on-shell Higgs, the complementary case of producing heavier states via an off-shell Higgs is relatively unexplored. In this paper we have commenced the systematic study of sensitivity to Higgs Portal states above threshold at both the LHC and potential future 100 TeV colliders. We have considered optimized searches in a variety of associated production modes, including vector boson fusion, gluon associated production, and $t \bar t$ associated production. We have taken particular care to correctly treat the effects of departure from the EFT limit in gluon associated production by appropriately re-weighting the results of leading-order HEFT Monte Carlo simulation.

Although the reach at 14 TeV is necessarily limited, there is nonetheless sufficient sensitivity to warrant optimized searches for heavy Higgs Portal states at Run 2. Searches at 100 TeV have the potential to substantially explore the Higgs Portal in regions of parameter space strongly motivated by physics beyond the Standard Model. In particular, regions motivated by dark matter, electroweak baryogenesis, and neutral naturalness can be effectively covered by searches for off-shell associated production of the Higgs. At the level of our analysis, the most promising channel appears to be vector boson fusion, but gluon- and $t \bar t$-associated production may also contribute substantial significance. The performance of mono-Higgs searches at 100 TeV, which we have omitted here, warrants further study. Searches in these channels also have the potential to directly discover or exclude Higgs Portal explanations of possible deviations in precision Higgs coupling measurements.

\section*{Acknowledgments}

We thank Nima Arkani-Hamed, Adam Falkowski, Jessie Shelton, Liantao Wang, and Andreas Weiler for helpful discussions. We particularly thank David Curtin, Patrick Meade, and Tien-Tien Yu for detailed discussion of the results in \cite{Curtin:2014jma}, and David Curtin and Prashant Saraswat for pointing out a numerical error in Section 2. A.T. acknowledges support from DOE grant DOE-SC0010008.
\appendix
\section{Electroweak Baryogenesis} \label{app:ewbg}

In this Appendix we sketch the details of our calculation of the viable parameter space for singlet-assisted electroweak baryogenesis. We work in the Lorentz gauge, where the gauge fixing parameter is $\xi=1$. The one-loop potential will include physical particles as well as goldstone bosons. The thermal potential is given by 
\begin{equation}
V(h, T) = V_{\rm tree} + V_{\rm CW} + V_{\rm thermal} + V_{\rm resum}
\end{equation}
where $V_{\rm tree}=-\mu^2h^2/2 + \lambda h^2/4$. $V_{\rm CW}$ is the one-loop zero temperature Coleman-Weinberg potential given by
\begin{equation}
V_{\rm CW}= \sum_i\frac{(-1)^F}{64\pi^2}m^4_i(h)
\left(\log\frac{m_i^2(h)}{m_i^2(v)}-\frac{3}{2}+\frac{2m_i^2(v)}{m_i^2(h)}\right)
\end{equation}
where the sum is over all degrees of freedom. The thermal potential is given by
\begin{equation}
V_{\rm thermal}= T^4\sum_i J_{F,B}(m_i(h)/T)= T^4\sum_i  \frac{(-1)^F}{2\pi^2}\int_0^\infty
\! dx \,x^2 \log \left[
1-(-1)^Fe^{\sqrt{x^2+m_i^2(h)/T^2}}\,
\right]
\end{equation}
In practical calculations, we expand the thermal integral $J_{F,B}(x)$ up to order $x^6$ in the small $x$ limit matched to bessel functions at large $x$. At larger temperature, there are infrared divergences that need to be resummed. The leading order resummed term is given by
\begin{equation}
V_{\rm resum}(h, T) = 
\sum_{i\in {\rm Boson}} \frac{T}{2\pi} {\rm Tr}\left[
m_i^3(h)+\Pi_i^{3/2} - (m_i^2(h)+\Pi_i)^{3/2}
\right]
\end{equation}
where $\Pi_i$ are the thermal masses. They are non-zero only for the higgs boson, goldstone, longitudinal gauge bosons and the $\phi$ scalar. The $\Pi_i$ are given by
\begin{eqnarray}
\Pi_h = \Pi_{\rm Goldstone} = T^2 \left(
\frac{3}{16}g^2 + \frac{1}{16}g'^2 + \frac{1}{4}y_t^2
+ \frac{1}{2}\lambda + \frac{1}{12}c_\phi
 \right) \\
\Pi_S = \frac{T^2}{3}c_\phi \notag \qquad
\Pi_{W_L}=T^2\frac{11}{6}g^2 \qquad
\Pi_{B_L}=T^2\frac{11}{6}g'^2 \, .
\end{eqnarray}

\bibliography{portalbib}
\bibliographystyle{jhep}

\end{document}